\documentclass[traditabstract]{aa} 
\usepackage{graphicx}
\usepackage{natbib}
\bibpunct{(}{)}{;}{a}{}{,}
\usepackage{epstopdf}
\usepackage{color}
\usepackage{rotating}
\usepackage{pdflscape}
\usepackage{booktabs}
\usepackage[varg]{txfonts}
\usepackage{hyperref}
\newcommand{\teff}{$T_{\rm eff}$}
\newcommand{\feh}{[M/H]}
\newcommand{\mh}{[M/H]}
\newcommand{\logg}{$\log g$}
\newcommand{\mlsep}{$\langle \Delta \nu \rangle$}
\newcommand{\numax}{$\nu_{\rm max}$}

\newcommand{\msol}{M$_{\odot}$}
\newcommand{\rsol}{R$_{\odot}$}
\newcommand{\lsol}{L$_{\odot}$}

\newcommand{\mhz}{$\mu$Hz}

\newcommand{\chisq}{$\chi^2$}

\newcommand{\yi}{$Y_i$}
\newcommand{\zi}{$Z_i$}

\usepackage{amstext}

\begin{document}

\title{Characterizing solar-type stars from full-length \emph{Kepler} data sets using the Asteroseismic Modeling Portal}

\author{O.~L.~Creevey\inst{1},
        T.~S.~Metcalfe\inst{2,3}, 
        M.~Schultheis\inst{1},
        D.~Salabert\inst{4},\\
        M.~Bazot\inst{5},
        F.~Th\'evenin\inst{1}, 
        S.~Mathur\inst{2},
        H.~Xu\inst{6},
        R.~A.~Garc\'ia\inst{4} }

\institute{
Universit\'e C\^ote d'Azur, Observatoire de la C\^ote d'Azur, CNRS, Laboratoire Lagrange, Bd de l'Observatoire, CS 34229, 06304 Nice cedex 4, France \email{orlagh.creevey@oca.eu}
\and 
Space Science Institute, 4750 Walnut St.\ Suite 205, Boulder CO 80301, USA
\and
Visiting Scientist, National Solar Observatory, 3665 Discovery Dr., Boulder CO 80303, USA
\and
Laboratoire AIM, CEA/DRF-CNRS, Universit\'e Paris 7 Diderot, IRFU/SAp, Centre de Saclay, 91191 Gif-sur-Yvette, France
\and
Center for Space Science, NYUAD Institute, New York University Abu Dhabi, PO Box 129188, Abu Dhabi, UAE
\and 
Computational \& Information Systems Laboratory, NCAR, P.O. Box 3000, Boulder CO 80307, USA
}

\date{Received ; accepted}

\abstract{The {\it Kepler} space telescope yielded unprecedented data for the 
study of solar-like oscillations in other stars. The large samples of 
multi-year observations posed an enormous data analysis challenge that has 
only recently been surmounted. Asteroseismic modeling has become more
sophisticated over time, with better methods gradually developing
alongside the extended observations and improved data analysis techniques. 
We apply the latest version of the Asteroseismic Modeling Portal (AMP) to the 
full-length {\it Kepler} data sets for 57 stars, comprising 
planetary hosts, binaries, solar-analogs, active stars, 
{and for validation purposes, the Sun}. 
From an analysis of the derived stellar properties for the full sample, 
we identify a variation of the mixing-length parameter with atmospheric properties.
We also derive a linear relation between the stellar age and 
a characteristic frequency separation ratio.
In addition, we find that the empirical correction for 
surface effects suggested by Kjeldsen and coworkers is
adequate for solar-type stars that are not much hotter (\teff\ $\lesssim 6200$~K) 
or significantly more evolved (\logg\ $\gtrsim 4.2$, \mlsep\ $\gtrsim 80\mu$Hz) 
than the Sun. 
Precise parallaxes from the Gaia mission and future observations 
from TESS and PLATO promise to improve the reliability of stellar properties 
derived from asteroseismology.}


\authorrunning{Creevey et al.}
\titlerunning{Characterizing solar-type stars}

    \maketitle

\section{Introduction\label{sec1}}

\begin{table*}
\begin{center}\caption{Spectroscopic constraints and complementary data of the {\it Kepler} targets \label{tab1}}
\begin{tabular}{lccccccccccccccccccc}
\hline\hline
KIC ID & \teff\  & \feh & $K_s$ & $A_{K_S}$ & P$_{\rm ROT}$ & Ref. \\
    & (K) & (dex) & (mag) & (mag) & (days) & \\
\hline
1435467&6326 $\pm$ 77&$+$0.01 $\pm$ 0.10&7.718 $\pm$ 0.009&0.011 $\pm$ 0.004&6.68 $\pm$ 0.89&1,A\\
2837475&6614 $\pm$ 77&$+$0.01 $\pm$ 0.10&7.464 $\pm$ 0.023&0.008 $\pm$ 0.002&3.68 $\pm$ 0.36&1,A\\
3427720&6045 $\pm$ 77&$-$0.06 $\pm$ 0.10&7.826 $\pm$ 0.009&0.020 $\pm$ 0.019&13.94 $\pm$ 2.15&1,B\\
3656476&5668 $\pm$ 77&$+$0.25 $\pm$ 0.10&8.008 $\pm$ 0.014&0.022 $\pm$ 0.050&31.67 $\pm$ 3.53&1,A\\
3735871&6107 $\pm$ 77&$-$0.04 $\pm$ 0.10&8.477 $\pm$ 0.016&0.018 $\pm$ 0.027&11.53 $\pm$ 1.24&1,A\\
4914923&5805 $\pm$ 77&$+$0.08 $\pm$ 0.10&7.935 $\pm$ 0.017&0.017 $\pm$ 0.029&20.49 $\pm$ 2.82&1,A\\
5184732&5846 $\pm$ 77&$+$0.36 $\pm$ 0.10&6.821 $\pm$ 0.005&0.012 $\pm$ 0.007&19.79 $\pm$ 2.43&1,A\\
5950854&5853 $\pm$ 77&$-$0.23 $\pm$ 0.10&9.547 $\pm$ 0.017&0.002 $\pm$ 0.004&&1\\
6106415&6037 $\pm$ 77&$-$0.04 $\pm$ 0.10&5.829 $\pm$ 0.017&0.003 $\pm$ 0.020&&1\\
6116048&6033 $\pm$ 77&$-$0.23 $\pm$ 0.10&7.121 $\pm$ 0.009&0.013 $\pm$ 0.020&17.26 $\pm$ 1.96&1,A\\
6225718&6313 $\pm$ 76&$-$0.07 $\pm$ 0.10&6.283 $\pm$ 0.011&0.003 $\pm$ 0.001&&1\\
6603624&5674 $\pm$ 77&$+$0.28 $\pm$ 0.10&7.566 $\pm$ 0.019&0.008 $\pm$ 0.008&&1\\
6933899&5832 $\pm$ 77&$-$0.01 $\pm$ 0.10&8.171 $\pm$ 0.015&0.023 $\pm$ 0.017&&1\\
7103006&6344 $\pm$ 77&$+$0.02 $\pm$ 0.10&7.702 $\pm$ 0.015&0.007 $\pm$ 0.010&4.62 $\pm$ 0.48&1,A\\
7106245&6068 $\pm$ 102&$-$0.99 $\pm$ 0.19&9.419 $\pm$ 0.006&0.015 $\pm$ 0.029&&4\\
7206837&6305 $\pm$ 77&$+$0.10 $\pm$ 0.10&8.575 $\pm$ 0.011&0.004 $\pm$ 0.005&4.04 $\pm$ 0.28&1,A\\
7296438&5775 $\pm$ 77&$+$0.19 $\pm$ 0.10&8.645 $\pm$ 0.009&0.012 $\pm$ 0.018&25.16 $\pm$ 2.78&1,A\\
7510397&6171 $\pm$ 77&$-$0.21 $\pm$ 0.10&6.544 $\pm$ 0.009&0.018 $\pm$ 0.010&&1\\
7680114&5811 $\pm$ 77&$+$0.05 $\pm$ 0.10&8.673 $\pm$ 0.006&0.011 $\pm$ 0.013&26.31 $\pm$ 1.86&1,A\\
7771282&6248 $\pm$ 77&$-$0.02 $\pm$ 0.10&9.532 $\pm$ 0.010&0.005 $\pm$ 0.001&11.88 $\pm$ 0.91&1,A\\
7871531&5501 $\pm$ 77&$-$0.26 $\pm$ 0.10&7.516 $\pm$ 0.017&0.023 $\pm$ 0.021&33.72 $\pm$ 2.60&1,A\\
7940546&6235 $\pm$ 77&$-$0.20 $\pm$ 0.10&6.174 $\pm$ 0.011&0.023 $\pm$ 0.009&11.36 $\pm$ 0.95&1,A\\
7970740&5309 $\pm$ 77&$-$0.54 $\pm$ 0.10&6.085 $\pm$ 0.011&0.003 $\pm$ 0.013&17.97 $\pm$ 3.09&1,A\\
8006161&5488 $\pm$ 77&$+$0.34 $\pm$ 0.10&5.670 $\pm$ 0.015&0.009 $\pm$ 0.006&29.79 $\pm$ 3.09&1,A\\
8150065&6173 $\pm$ 101&$-$0.13 $\pm$ 0.15&9.457 $\pm$ 0.014&0.010 $\pm$ 0.013&&4\\
8179536&6343 $\pm$ 77&$-$0.03 $\pm$ 0.10&8.278 $\pm$ 0.009&0.005 $\pm$ 0.016&24.55 $\pm$ 1.61&1,A\\
8379927&6067 $\pm$ 120&$-$0.10 $\pm$ 0.15&5.624 $\pm$ 0.011&0.004 $\pm$ 0.012&16.99 $\pm$ 1.35&2,A\\
8394589&6143 $\pm$ 77&$-$0.29 $\pm$ 0.10&8.226 $\pm$ 0.016&0.013 $\pm$ 0.010&&1\\
8424992&5719 $\pm$ 77&$-$0.12 $\pm$ 0.10&8.843 $\pm$ 0.011&0.016 $\pm$ 0.018&&1\\
8694723&6246 $\pm$ 77&$-$0.42 $\pm$ 0.10&7.663 $\pm$ 0.007&0.003 $\pm$ 0.001&&1\\
8760414&5873 $\pm$ 77&$-$0.92 $\pm$ 0.10&8.173 $\pm$ 0.009&0.016 $\pm$ 0.012&&1\\
8938364&5677 $\pm$ 77&$-$0.13 $\pm$ 0.10&8.636 $\pm$ 0.016&0.003 $\pm$ 0.009&&1\\
9025370&5270 $\pm$ 180&$-$0.12 $\pm$ 0.18&7.372 $\pm$ 0.025&0.041 $\pm$ 0.030&&3\\
9098294&5852 $\pm$ 77&$-$0.18 $\pm$ 0.10&8.364 $\pm$ 0.009&0.011 $\pm$ 0.021&19.79 $\pm$ 1.33&1,A\\
9139151&6302 $\pm$ 77&$+$0.10 $\pm$ 0.10&7.952 $\pm$ 0.014&0.002 $\pm$ 0.011&10.96 $\pm$ 2.22&1,B\\
9139163&6400 $\pm$ 84&$+$0.15 $\pm$ 0.09&7.231 $\pm$ 0.007&0.013 $\pm$ 0.007&&6\\
9206432&6538 $\pm$ 77&$+$0.16 $\pm$ 0.10&8.067 $\pm$ 0.013&0.032 $\pm$ 0.037&8.80 $\pm$ 1.06&1,A\\
9353712&6278 $\pm$ 77&$-$0.05 $\pm$ 0.10&9.607 $\pm$ 0.011&0.011 $\pm$ 0.010&11.30 $\pm$ 1.12&1,A\\
9410862&6047 $\pm$ 77&$-$0.31 $\pm$ 0.10&9.375 $\pm$ 0.013&0.011 $\pm$ 0.001&22.77 $\pm$ 2.37&1,A\\
9414417&6253 $\pm$ 75&$-$0.13 $\pm$ 0.10&8.407 $\pm$ 0.009&0.010 $\pm$ 0.010&10.68 $\pm$ 0.66&7,A\\
9955598&5457 $\pm$ 77&$+$0.05 $\pm$ 0.10&7.768 $\pm$ 0.017&0.002 $\pm$ 0.001&34.20 $\pm$ 5.64&1,A\\
9965715&5860 $\pm$ 180&$-$0.44 $\pm$ 0.18&7.873 $\pm$ 0.012&0.005 $\pm$ 0.005&&3\\
10079226&5949 $\pm$ 77&$+$0.11 $\pm$ 0.10&8.714 $\pm$ 0.012&0.015 $\pm$ 0.025&14.81 $\pm$ 1.23&1,A\\
10454113&6177 $\pm$ 77&$-$0.07 $\pm$ 0.10&7.291 $\pm$ 9.995&0.042 $\pm$ 0.019&14.61 $\pm$ 1.09&1,A\\
10516096&5964 $\pm$ 77&$-$0.11 $\pm$ 0.10&8.129 $\pm$ 0.015&0.000 $\pm$ 0.012&&1\\
10644253&6045 $\pm$ 77&$+$0.06 $\pm$ 0.10&7.874 $\pm$ 0.021&0.008 $\pm$ 0.015&10.91 $\pm$ 0.87&1,A\\
10730618&6150 $\pm$ 180&$-$0.11 $\pm$ 0.18&7.874 $\pm$ 0.021&0.008 $\pm$ 0.015&&3\\
10963065&6140 $\pm$ 77&$-$0.19 $\pm$ 0.10&7.486 $\pm$ 0.011&0.003 $\pm$ 0.016&12.58 $\pm$ 1.70&1,A\\
11081729&6548 $\pm$ 82&$+$0.11 $\pm$ 0.10&7.973 $\pm$ 0.011&0.005 $\pm$ 0.001&2.74 $\pm$ 0.31&1,A\\
11253226&6642 $\pm$ 77&$-$0.08 $\pm$ 0.10&7.459 $\pm$ 0.007&0.017 $\pm$ 0.013&3.64 $\pm$ 0.37&1,A\\
11772920&5180 $\pm$ 180&$-$0.09 $\pm$ 0.18&7.981 $\pm$ 0.014&0.008 $\pm$ 0.005&&3\\
12009504&6179 $\pm$ 77&$-$0.08 $\pm$ 0.10&8.069 $\pm$ 0.019&0.005 $\pm$ 0.034&9.39 $\pm$ 0.68&1,A\\
12069127&6276 $\pm$ 77&$+$0.08 $\pm$ 0.10&9.494 $\pm$ 0.012&0.016 $\pm$ 0.005&0.92 $\pm$ 0.05&1,A\\
12069424&5825 $\pm$ 50&$+$0.10 $\pm$ 0.03&4.426 $\pm$ 0.009&0.005 $\pm$ 0.006&23.80 $\pm$ 1.80&5,B\\
12069449&5750 $\pm$ 50&$+$0.05 $\pm$ 0.02&4.651 $\pm$ 0.005&0.005 $\pm$ 0.006&23.20 $\pm$ 6.00&5,B\\
12258514&5964 $\pm$ 77&$+$-0.00 $\pm$ 0.10&6.758 $\pm$ 0.011&0.021 $\pm$ 0.021&15.00 $\pm$ 1.84&1,A\\
12317678&6580 $\pm$ 77&$-$0.28 $\pm$ 0.10&7.631 $\pm$ 0.009&0.027 $\pm$ 0.021&&1\\
\hline\hline
\end{tabular}
\end{center}
Spectroscopic references: $^1$\cite{Buchhave2015}, $^2$\cite{Ramirez2009}, $^3$\cite{Pinsonneault2012}, $^4$\cite{Huber2013}, $^5$\cite{Chaplin2014}, $^6$\cite{Pinsonneault2014},$^7$\cite{Casagrande2014}\\
Rotation period references: $^{\rm A}$\cite{garcia2014}, $^{\rm B}$\cite{cellier2016}
\end{table*}

\begin{figure*}
\center{\includegraphics[width=0.98\textwidth]{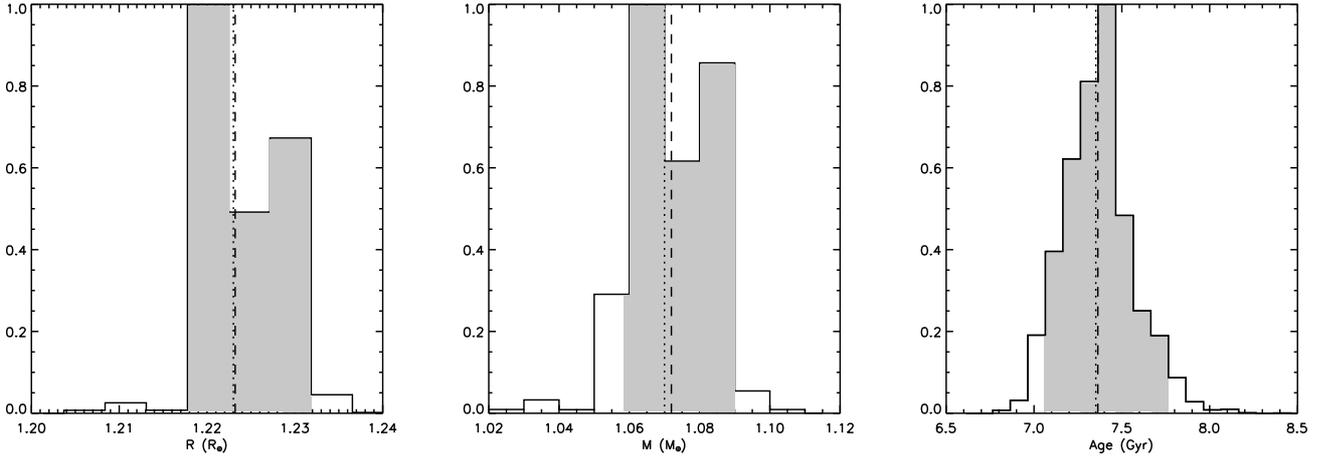}}
\caption{Normalized posterior probability functions for KIC~12069424.  
From left to right we show radius, mass, and age. We also show the adopted 
parameter $\langle P \rangle$ (dashed line), 
the 68\% region (shaded) and $P_{\rm AMP}$ (dotted line). 
\label{fig:examplelikelihood}}
\end{figure*}

Solar-like oscillations are stochastically excited and intrinsically
damped by turbulent motions in the near-surface layers of stars with
substantial outer convection zones. The sound waves produced by these
motions travel through the interior of the star, and those with resonant
frequencies drive global oscillations that modulate the integrated
brightness of the star by a few parts per million and change the 
{surface} radial velocity by several meters per second. The characteristic
timescale of these variations is determined by the sound
travel time across the stellar diameter, which is around 5 minutes for a
star like the Sun. With sufficient precision, more than a 
dozen consecutive overtones can be detected for each set of oscillation 
modes with radial, dipole, quadrupole, and sometimes even octupole
geometry {(i.e., for $l=0, 1, 2,$ and 3, respectively, where $l$ is
the angular degree)}. The technique of asteroseismology uses these 
oscillation frequencies {combined} with other observational constraints 
to measure the stellar radius, mass, age, and other properties of the 
stellar interior \citep[for a recent review, see][]{ChaplinMiglio2013}.

The {\it Kepler} space telescope yielded unprecedented data for the
study of solar-like oscillations in other stars. Ground-based radial
velocity data had previously allowed the detection of solar-like
oscillations in some of the brightest stars in the sky
\citep[e.g.,][]{Brown1991, Kjeldsen1995, Bedding2001, Bouchy2002,
Carrier2003}, but {intensive} multi-site campaigns were required to measure
and identify the frequencies unambiguously \cite[e.g.,][]{Arentoft2008}.
The {\it Convection Rotation and planetary Transits} satellite (CoRoT, 
\citealt{Baglin2006}) achieved the {photometric} precision
necessary to detect solar-like oscillations in main-sequence stars
\citep[e.g.,][]{Michel2008}, and it obtained continuous photometry for up
to five months. NASA's {\it Kepler} mission \citep{Borucki2010} extended 
these initial successes to a larger sample of solar-type stars, with 
observations eventually spanning up to several years \citep{Chaplin2010}. 
Precise photometry from {\it Kepler} led to the detection of solar-like 
oscillations in nearly 600 main-sequence and subgiant stars 
\citep{Chaplin2014}, including the measurement of individual frequencies 
in more than 150 targets \citep{Appourchaux2012, Davies2016, Lund2016}.

Asteroseismic modeling has become more sophisticated over time, with
better methods gradually developing alongside the {extended} observations
and {improved} data analysis techniques.
Initial efforts attempted to reproduce the observed large and small
frequency separations with models that simultaneously matched constraints
from spectroscopy \citep[e.g.,][]{JCD1995, Thevenin2002, Fernandes2003,
Thoul2003}. As individual oscillation frequencies became available,
modelers started to match the observations in \'echelle diagrams that
{highlighted variations around the average frequency separations}
\citep[e.g.,][]{DiMauro2003, Guenther2004, Eggenberger2004}. This approach
continued until the frequency precision from longer space-based
observations became sufficient to reveal systematic errors in the models
that are known as {\it \textit{\textup{surface effects}}}, which arise from incomplete modeling of
the near-surface layers where the mixing-length treatment of convection
is {approximate}. \cite{Kjeldsen2008} proposed an empirical correction for {the} surface effects based on the 
discrepancy for {the} standard solar model, 
and applied it to ground-based observations of several stars with different 
masses and evolutionary states. The correction was subsequently implemented 
using stars observed by {CoRoT} and {\it Kepler} \citep{Kallinger2010,
Metcalfe2010}.

During the {\it Kepler} mission, asteroseismic modeling methods were
adapted as longer data sets became available. The first year of
short-cadence data \citep[{sampled at} 58.85~s,][]{Gilliland2010} 
was devoted to an
asteroseismic survey of 2000 solar-type stars observed for one
month each.
The survey initially yielded {frequencies for} 22 stars, {allowing} 
detailed modeling \citep{Mathur2012}, and hundreds of
targets were flagged for extended observations during the remainder of
the mission. Longer data sets improved the signal-to-noise ratio (S/N)
{of the power spectrum} for stars with {previously} marginal detections, and 
yielded additional oscillation frequencies for the best targets in the sample.
The first coordinated analysis of nine-month data sets yielded individual   
frequencies in 61 stars \citep{Appourchaux2012}, though many were 
subgiants with complex patterns of dipole mixed-modes. The larger   
set of radial orders observed in each star began to reveal the
limitations of the empirical correction for surface effects
\citep{Metcalfe2014}. This situation {motivated the implementation} 
of a Bayesian method that marginalized over the unknown systematic error 
{for each frequency} \citep{Gruberbauer2012}, as well as a
method for fitting ratios of frequency separations that are insensitive to
surface effects \citep{Roxburgh2003, Bazot2013, SilvaAguirre2013}. It also inspired
the development of a more physically motivated correction {based on
an analysis of frequency shifts induced by the solar magnetic cycle}
{\citep{Gough1990, Ball2014, schmittbasu2015}}. 
The {\it Kepler} telescope completed its primary 
mission in 2013, but the large samples of multi-year observations posed 
an enormous data analysis challenge that has only recently been surmounted 
\citep{benomar2014a, benomar2014b, Davies2015, Davies2016, Lund2016}. The first modeling 
of these full-length data sets appeared in \cite{SilvaAguirre2015} and 
\cite{Metcalfe2015}.

In this paper we apply the latest version of the Asteroseismic Modeling
Portal \citep[hereafter AMP, see][]{Metcalfe2009} to {oscillation 
frequencies derived from} the full-length {\it Kepler} {observations}
for 57 stars, {as determined by \cite{Lund2016}}. The new fitting 
method relies on ratios of frequency  
separations rather than the individual frequencies, so that we can use the
modeling results to investigate the empirical amplitude and character of
{the} surface effects within the sample. 
We describe the sources of our adopted
observational constraints in Section~\ref{sec2}. We outline
updates to the AMP input physics and fitting methods in Section~\ref{sec3}, 
including an overview of how the optimal stellar properties and their 
uncertainties are determined. In Section~\ref{sec4} we present 
the modeling results, and in Section~\ref{sec:sec5} we use them to establish 
the limitations of the \cite{Kjeldsen2008} correction for surface 
effects. Finally, after summarizing in Section~\ref{sec6}, we discuss our expectations for 
asteroseismic modeling of future observations from the Transiting Exoplanet
Survey Satellite, \citep[TESS,][]{tess-ricker2015} and PLanetary Transits 
and Oscillations of stars \citep[PLATO,][]{plato-rauer2014} missions.

\section{Observational constraints\label{sec2}}

To constrain the properties of each star in our sample, we adopted the
solar-like oscillation frequencies determined by \cite{Lund2016} from a
uniform analysis of the full-length {\it Kepler} data sets. {For each target}, 
the power spectrum of the time-series photometry shows the oscillations
embedded in several background components attributed to granulation,
faculae, and shot noise. The {power spectral distribution of} 
individual {modes} were modeled as
Lorentzian functions, and the background components were optimized
simultaneously in a Bayesian manner using the procedure described in
\cite{Lund2014}. For the targets presented here, this analysis resulted in
sets of oscillation modes spanning 7 to 20 
radial orders. In most cases, the identified 
frequencies included only $l=0$, 1, and 2 modes, but {for} 14 stars,
{the mode-fitting procedure} also {identified} 
limited sets of $l=3$ modes spanning 2 to 6 radial orders. Complete
tables of the identified frequencies for each star are published in
\cite{Lund2016}.

To complement the {oscillation frequencies}, we also adopted spectroscopic
constraints on the effective temperature, $T_{\rm eff}$, and metallicity,
[M/H], for each star. For 46 of the targets in our sample, we used the
uniform spectroscopic analysis of \cite{Buchhave2015}. In this case, the
values and uncertainties on $T_{\rm eff}$ and [M/H] were determined using
the Stellar Parameters Classification (SPC) method 
described in detail by
\cite{Buchhave2012,Buchhave2014}. For the other 11 stars in our sample, 
{which were not included in \cite{Buchhave2015}}, we
adopted constraints from a variety of sources, including 
\cite{Ramirez2009,Pinsonneault2012,Huber2013,Chaplin2014,Pinsonneault2014}, 
and from the SAGA survey \citep{Casagrande2014}. The {57 stars in our sample} 
span a range of $T_{\rm eff}$ from 5180 to 6642~K and [M/H] from $-$0.99 to 0.36 dex.
These atmospheric constraints are listed in Table~\ref{tab1} along with {the
K-band magnitude from 2MASS, $K_s$ \citep{2MASS}, the derived interstellar
absorption, $A_{Ks}$ (see Section~\ref{sec:asteroseismicdistances}),} 
and rotational periods from \citet{garcia2014} and \citet{cellier2016}. 
Although independent determinations of the 
radius and luminosity are available for a few of the stars in our sample, we 
excluded these constraints from the modeling so that we could use them to 
assess the accuracy of our results (see Section~\ref{sec4}).

\begin{table}[]
\centering
\caption{Reference solar parameters from AMP using
the updated method and physics}
\begin{tabular}{lcccccc}
\hline\hline
 & AMP$_1$ &AMP$_2$ &AMP$_3$ &AMP$_4$ & $\langle P \rangle$ & $\sigma$  \\
\hline
$R$ (\rsol) & 1.002&1.003&1.003&1.010& 1.001 & 0.005 \\
$M$ (\msol) & 1.01 & 1.01 & 1.01 & 1.03& 1.001 & 0.02 \\
Age (Gyr) & 4.59 & 4.38 & 4.41 & 4.69 & 4.38 & 0.22 \\
\zi &  0.019 & 0.021 & 0.020 & 0.024&0.017 & 0.002 \\
\yi & 0.266 & 0.281 & 0.278 & 0.282& 0.265 & 0.023 \\
$\alpha$ & 2.16 & 2.24 & 2.24 & 2.30& 2.12 & 0.12 \\
$L$ (\lsol) & 0.96 &0.99&0.99&1.00& 0.97 & 0.03\\
$\log g$ (dex) &4.441 &4.439&4.439&4.442& 4.438 & 0.003\\
$\chi^2$ & 1.047 & 0.968 & 0.995 & 1.058\\
\hline\hline
\end{tabular}
\label{tab:solar-reference}
\end{table}

\begin{figure}
    \includegraphics[width=0.48\textwidth]{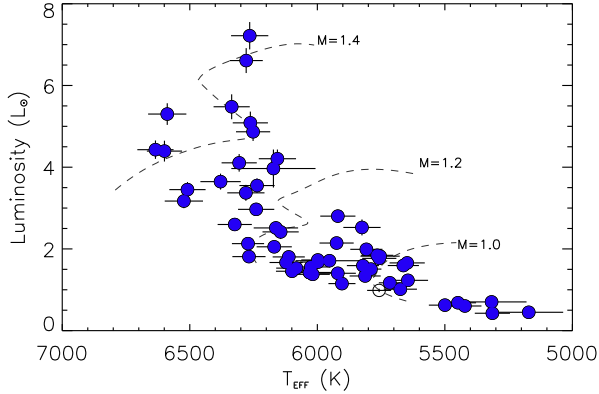}
    \caption{\label{fig:hrdiag}HR diagram showing the position of the sample of stars used for this work.  
    Evolutionary tracks for solar-metallicity models with 1.0, 1.2, and 1.4 \msol\ stellar masses are shown.}
\end{figure}

\section{Asteroseismic modeling\label{sec3}}

{Based on} the observational constraints described in Section~\ref{sec2},
{we determined} the properties of each star in our sample using the 
latest version of AMP. {The method} {relies on} a parallel genetic algorithm 
\citep[hereafter GA, see][]{Metcalfe2003} to optimize the match between the 
{properties of a stellar} model and a given set of observations. The
asteroseismic models {are} generated by the Aarhus stellar evolution and 
adiabatic pulsation codes \citep{JCD08a,JCD08b}. The
search procedure generates thousands of models that can be
used to evaluate the stellar properties and their uncertainties. Unlike 
the usual grid-modeling approach, the GA preferentially samples 
combinations of model parameters that provide a better than average match 
to the observations. This approach allows us not only to identify the 
globally optimal solution, but also to include the effects of parameter 
correlations and non-uniqueness into reliable uncertainties. Below we outline 
recent updates to the input physics and {model-fitting}  methods, and we 
describe improvements to our statistical analysis of the results.

\subsection{Updated physics and methods}

The AMP code has been in development since 2004. Details about previous 
versions are outlined in \cite{Metcalfe2015}. For this paper we use version 
1.3, which includes input physics that are mostly unchanged from version 1.2
\citep{Metcalfe2014}. It uses the {2005 release of the} OPAL equation of state 
\citep{Rogers2002}, with opacities from OPAL \citep{Iglesias1996} supplemented 
by \cite{Ferguson2005} at low temperatures. Nuclear reaction rates come from 
the NACRE collaboration \citep{Angulo1999}. The prescription of \cite{Michaud1993} 
for diffusion and settling is applied to helium, but not to heavier elements because some models are numerically instable. Convection is described using the 
mixing-length treatment of \cite{BohmVitense1958} with no overshoot.

There have been several minor updates to the model physics for version 1.3 of the 
AMP code. First, it incorporates the revised $^{14}N+p$ reaction from NACRE 
\citep{Angulo2005}, which is particularly important for more evolved stars. Second,
it uses the solar mixture of \cite{GS1998} instead of \cite{GN1993}. This requires 
different opacity tables and a slight modification to the calculation of metallicity 
[$\log(Z_\odot/X_\odot) = -1.64$ instead of $-$1.61]. Finally, following the suggestion 
of \cite{SilvaAguirre2015}, diffusion and settling is only applied to models with 
$M < 1.2\ M_\odot$, to avoid potential biases that are due to the short diffusion timescales 
in the envelopes of more massive stars.

\begin{figure*}
    \centering
                \includegraphics[width=0.45\textwidth]{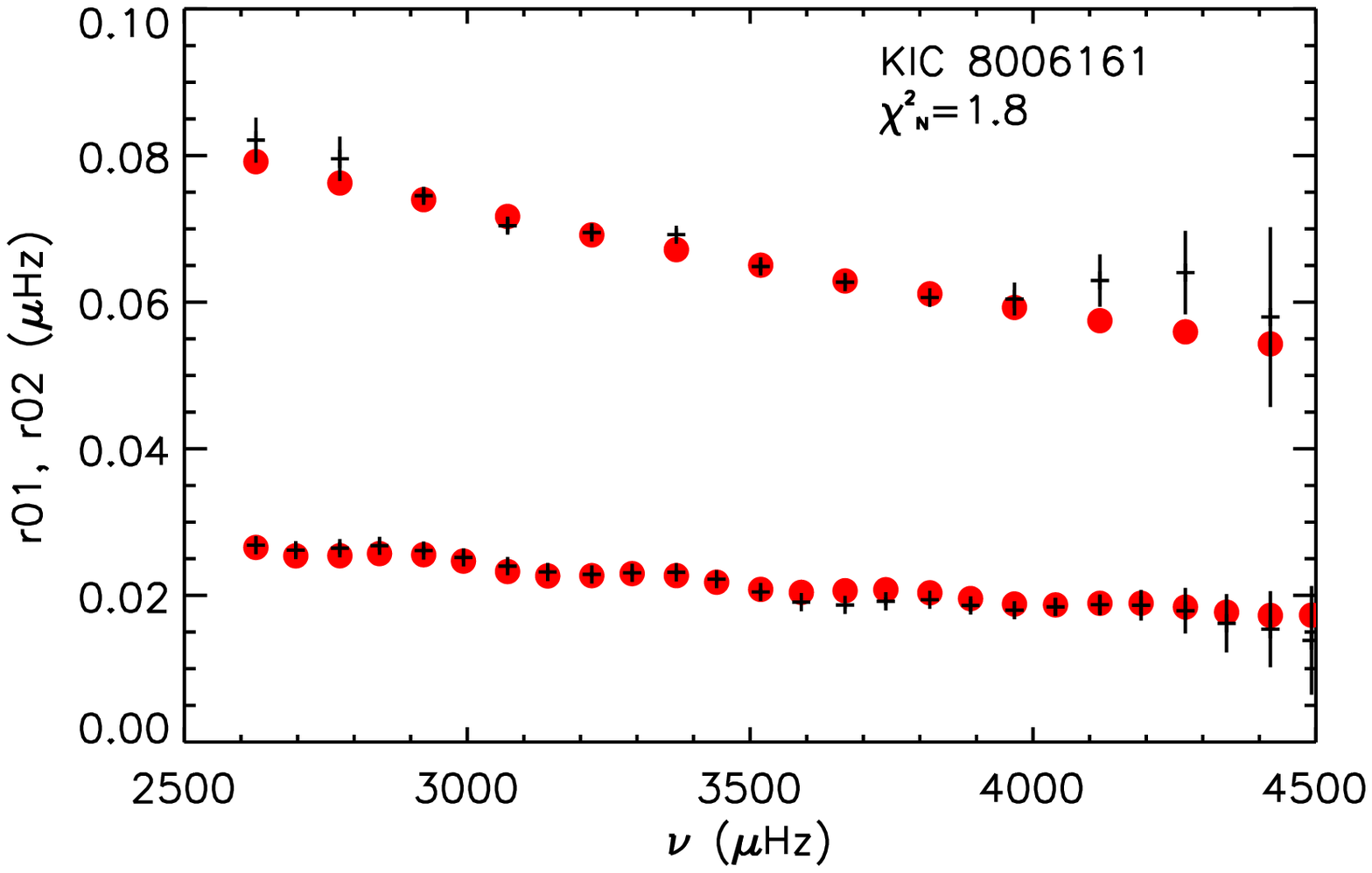}
            \includegraphics[width=0.45\textwidth]{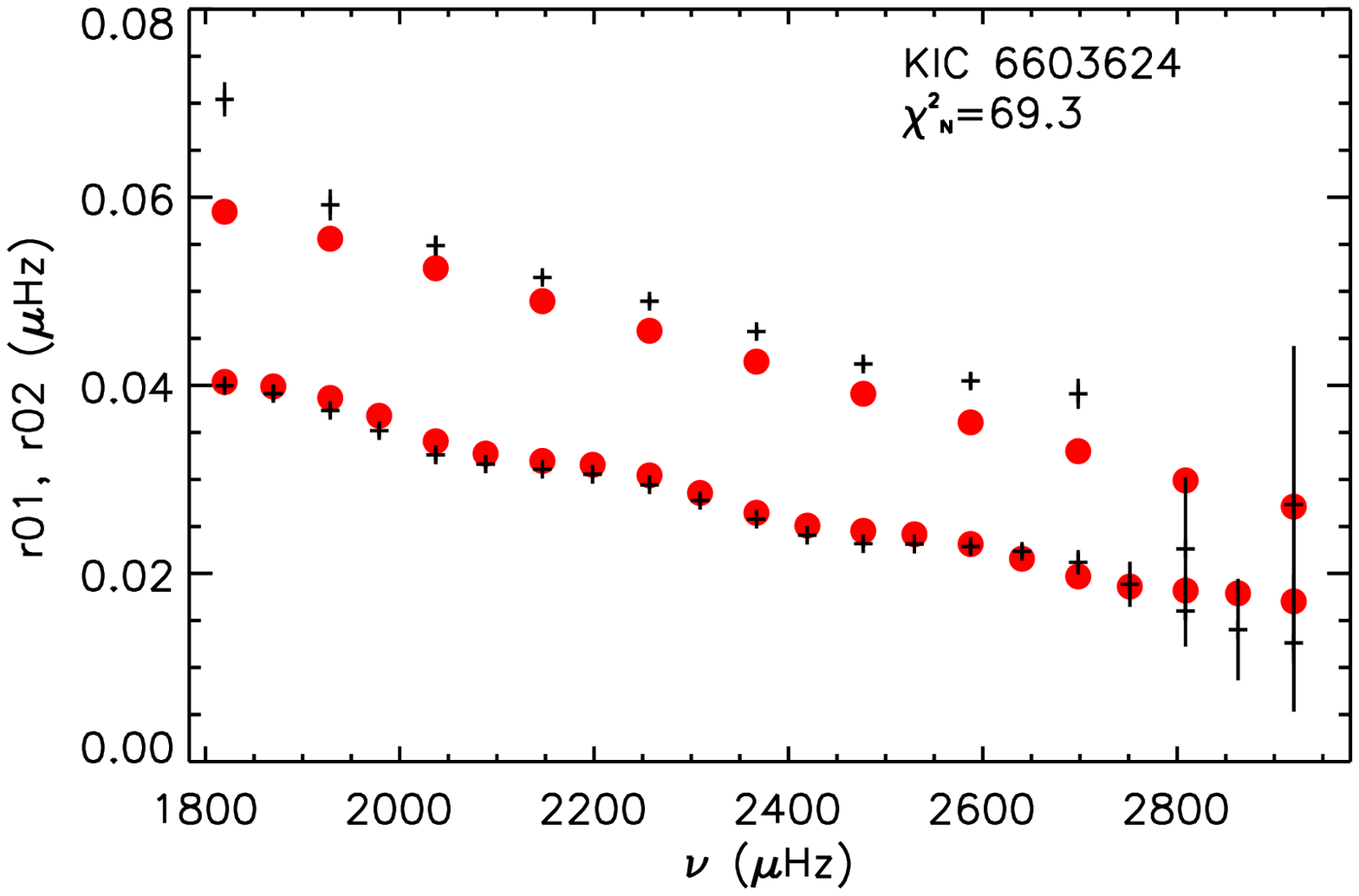}
    \includegraphics[width=0.45\textwidth]{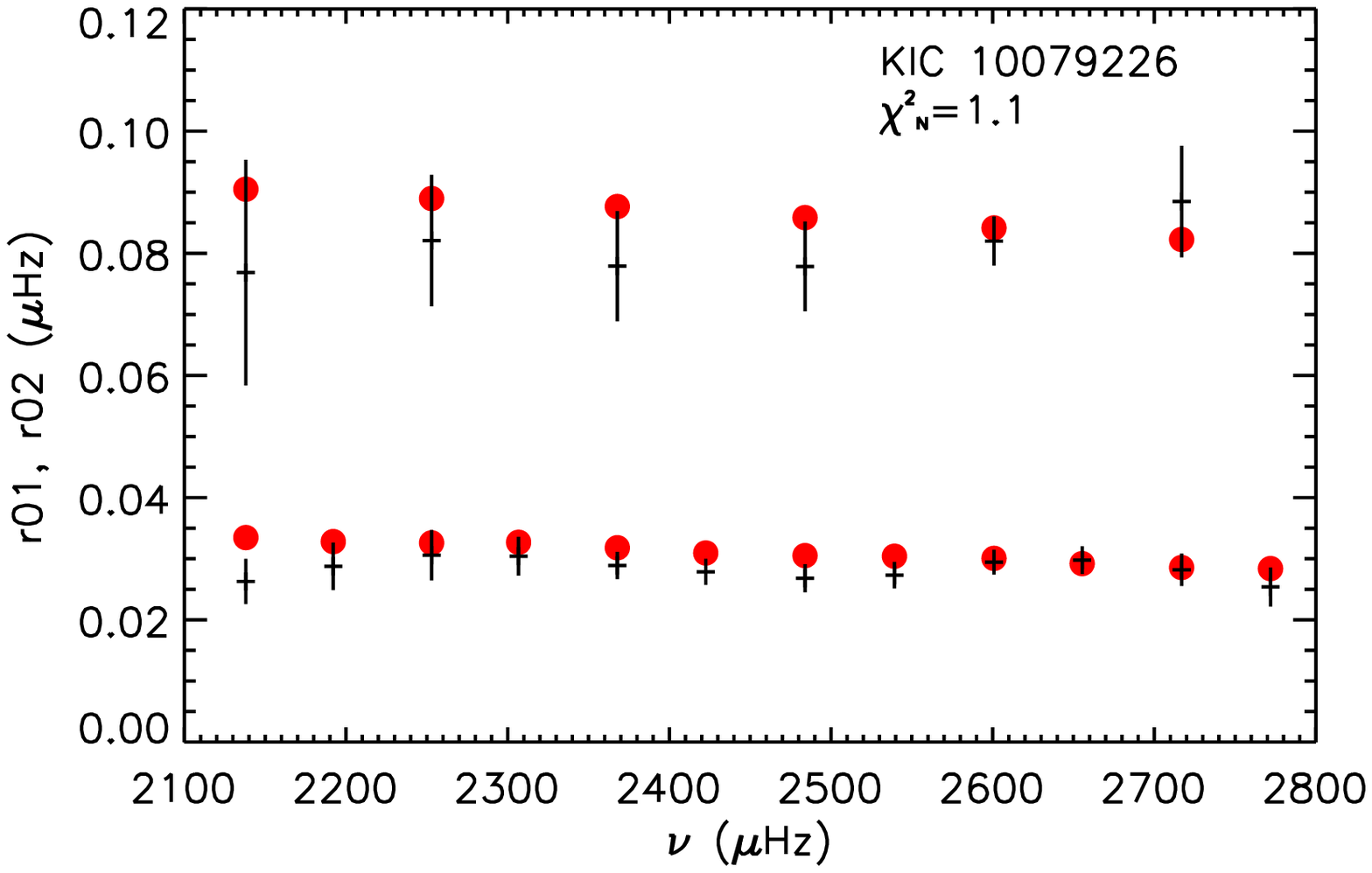}
        \includegraphics[width=0.45\textwidth]{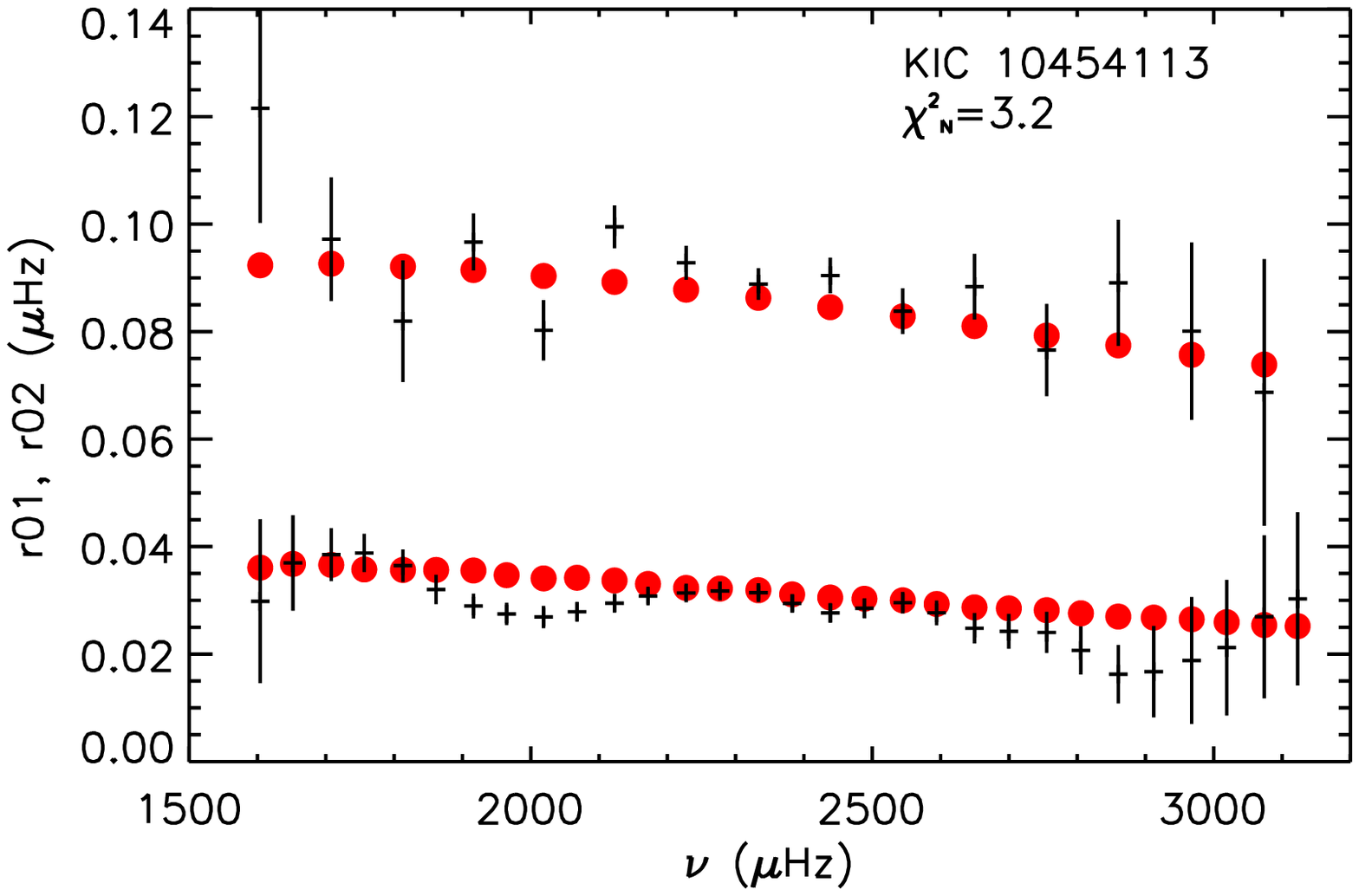}
    \caption{Representative examples of fits to the seismic 
    frequency ratios.}
    \label{fig:fitdata0}
\end{figure*}

{The frequency separation ratios $r_{01}$ and $r_{02}$ were defined by 
\cite{Roxburgh2003} as}
\begin{equation}
r_{01}(n) = \frac{\nu_{n-1,0} - 4\nu_{n-1,1} + 6\nu_{n,0} - 4\nu_{n,1} + \nu_{n+1,0}}{8(\nu_{n,1} - \nu_{n-1,1})}
\label{eqn:r01}
\end{equation}
and 
\begin{equation}
r_{02}(n) = \frac{\nu_{n,0} - \nu_{n-1,2}}{\nu_{n,1} - \nu_{n-1,1}},
\end{equation}
{where $\nu$ is the mode frequency, $n$ is the radial order, and $l$ is 
the angular degree.} These ratios were first included as observational 
constraints in AMP~1.2. Version 1.3 uses these ratios exclusively, omitting 
the individual oscillation frequencies to avoid potential biases from the 
empirical correction for surface effects. AMP~1.3 also {calculates} the 
full covariance matrix of $r_{01}$, which is necessary to properly account 
for correlations 
induced by the five-point smoothing that is 
{implicit 
in Eq.~(\ref{eqn:r01})}.

{For each stellar model, AMP~1.2 defined the quality of the match to 
observations using a combination of metrics from four different sets of
constraints. For AMP~1.3, we} combine all observational constraints into 
a single $\chi^2$ metric
\begin{equation}
\chi^2 = (x - x_M)^T C^{-1} (x - x_M),
\label{eqn:chisq}
\end{equation}
where $C$ is the covariance matrix of the observational constraints $x$, 
and $x_M$ are the corresponding observables from the model. For the results
presented here, $x$ includes only the ratios $r_{01}$ and $r_{02}$ 
{augmented by} the atmospheric constraints \teff\ and \mh. {$C$ is 
assumed to be diagonal for all observables except $r_{01}$. Like all 
previous versions of AMP, the individual frequencies are} used to 
calculate the average large separation of the radial modes $\Delta\nu_0$, 
allowing us to optimize the {stellar} age {along each model sequence}
and then match the lowest observed radial mode frequency \citep[see][]{Metcalfe2009}.

\subsection{Statistical analysis}

Versions {1.0 and 1.1} of the AMP code performed a local analysis 
near the optimal model to determine the uncertainties on each  
parameter \citep{Metcalfe2009, Mathur2012}. This approach failed to capture
the uncertainties due to parameter correlations and non-uniqueness of the
solution, so that it typically {produced} implausibly small error bars,
although formally correct. 
To {derive} more realistic uncertainties {in version 1.2}, 
\cite{Metcalfe2014} began using the thousands of models sampled by the GA {during the optimization procedure. As the GA approaches the optimal model, 
each parameter is densely sampled with a uniform spacing in stellar mass 
($M$), initial metallicity ($Z_i$), initial helium mass fraction ($Y_i$), 
and mixing-length ($\alpha$)}. Each sampled model is assigned a likelihood
 \begin{equation}
 \mathcal{L}=\exp\left( \frac{-\chi^2}{2} \right),
 \label{eqn:likelihooddefn}
 \end{equation}
where $\chi^2$ is calculated from Eq.~(\ref{eqn:chisq}). 
By assuming flat priors on each of the model parameters, we then 
construct posterior probability functions (PPF) for each of the
stellar properties to obtain more reliable estimates of the values and  
uncertainties from the dense ensemble of models sampled by the GA. We adopt 
the median value of the PPF as the best estimate for the parameter value, 
$\langle P \rangle$. We use the 68\% credible interval {of the PPF} to 
define the associated uncertainty, $\sigma$. Sample PPFs for the radius, 
mass, and age of KIC~12069424 are shown in Fig.~\ref{fig:examplelikelihood}.

Combining the best estimates for each of the stellar properties generally
will not produce the best stellar model. For many purposes it is useful
to identify a {\it \textup{reference model}}, an individual stellar model that is
representative of the PPF. The optimal model identified by 
AMP, $P_{\rm AMP}$, is used as the reference model, but it can sometimes fall 
near the edge of one or more of the distributions. 
A comparison of the masses and ages estimated from $\langle P \rangle$ 
and $P_{\rm AMP}$ yields differences much smaller than 1$\sigma$ for most cases.  

\subsection{Validation with solar data}

To validate our new approach, we used AMP~1.3 to {match a set of solar
oscillation frequencies comparable to the {\it Kepler} observations of
16~Cyg~A and B \citep{Metcalfe2015}. The frequencies were derived from
observations obtained with} the Variability of solar IRradiance and Gravity 
Oscillations (VIRGO) instrument \citep{virgo} {using 2.5 years of data \citep{Davies2015}}.
The best models identified by the four independent runs of the GA are listed 
in Table~\ref{tab:solar-reference} under the headings AMP$_N$ along with 
their individual $\chi^2$ values\footnote{\small{See \url{https://amp.phys.au.dk/browse/simulation/829}} {for details of that AMP modeling}}.
The model with the lowest value of $\chi^2$ is the optimal solution 
identified by AMP, and this is adopted as the reference model. The remaining 
models reveal intrinsic parameter correlations, in particular between the
mass and initial composition. The final two columns of Table~\ref{tab:solar-reference} 
show the values of $\langle P \rangle$ and $\sigma$ derived from the PPFs,
showing excellent agreement with the known solar properties: {$R, M, L \equiv 1$, 
age\,$=4.60\pm0.04$~Gyr \citep{Houdek2011}.}

\begin{figure}
    \centering
    \includegraphics[width=0.48\textwidth]{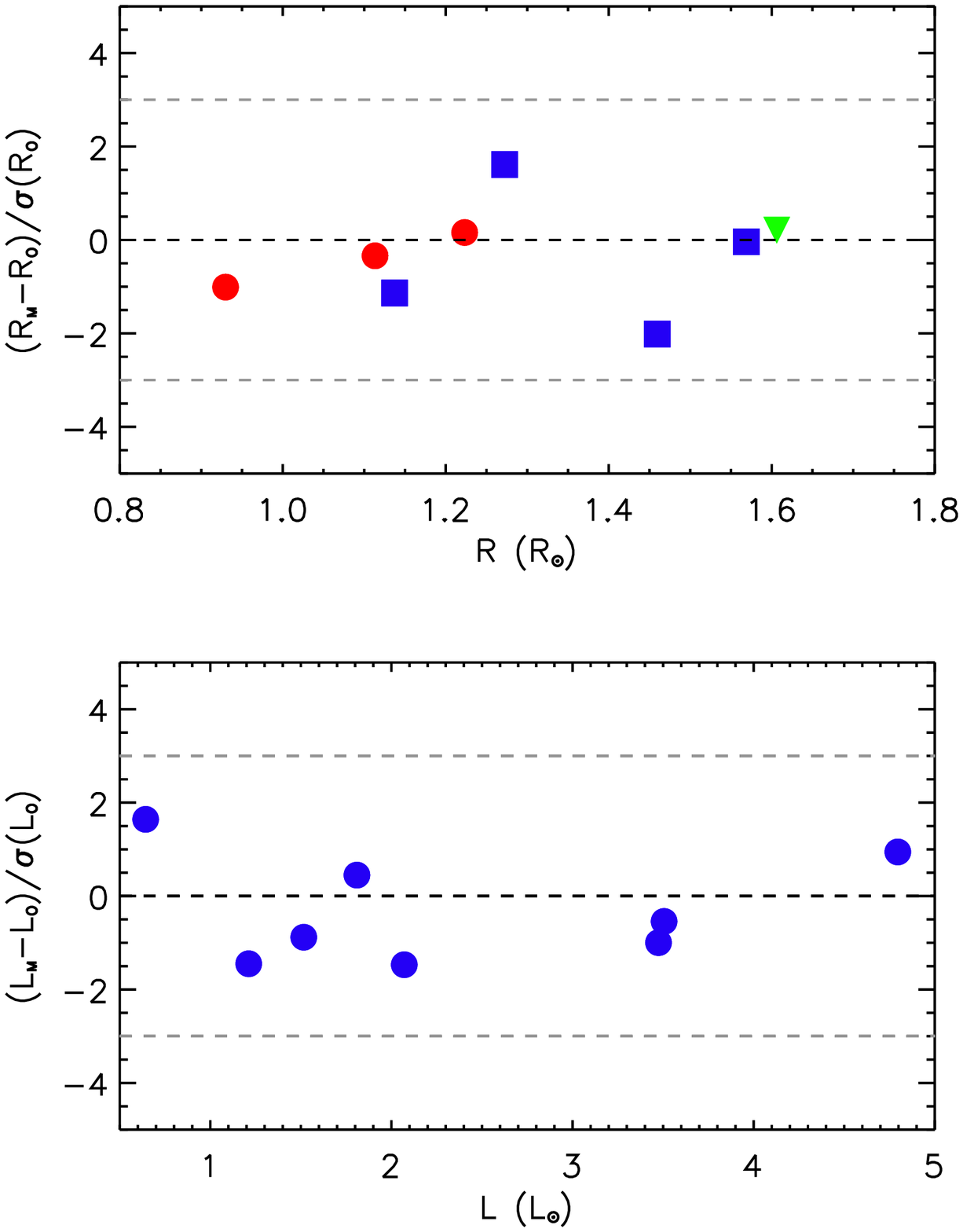}
    \includegraphics[width=0.48\textwidth]{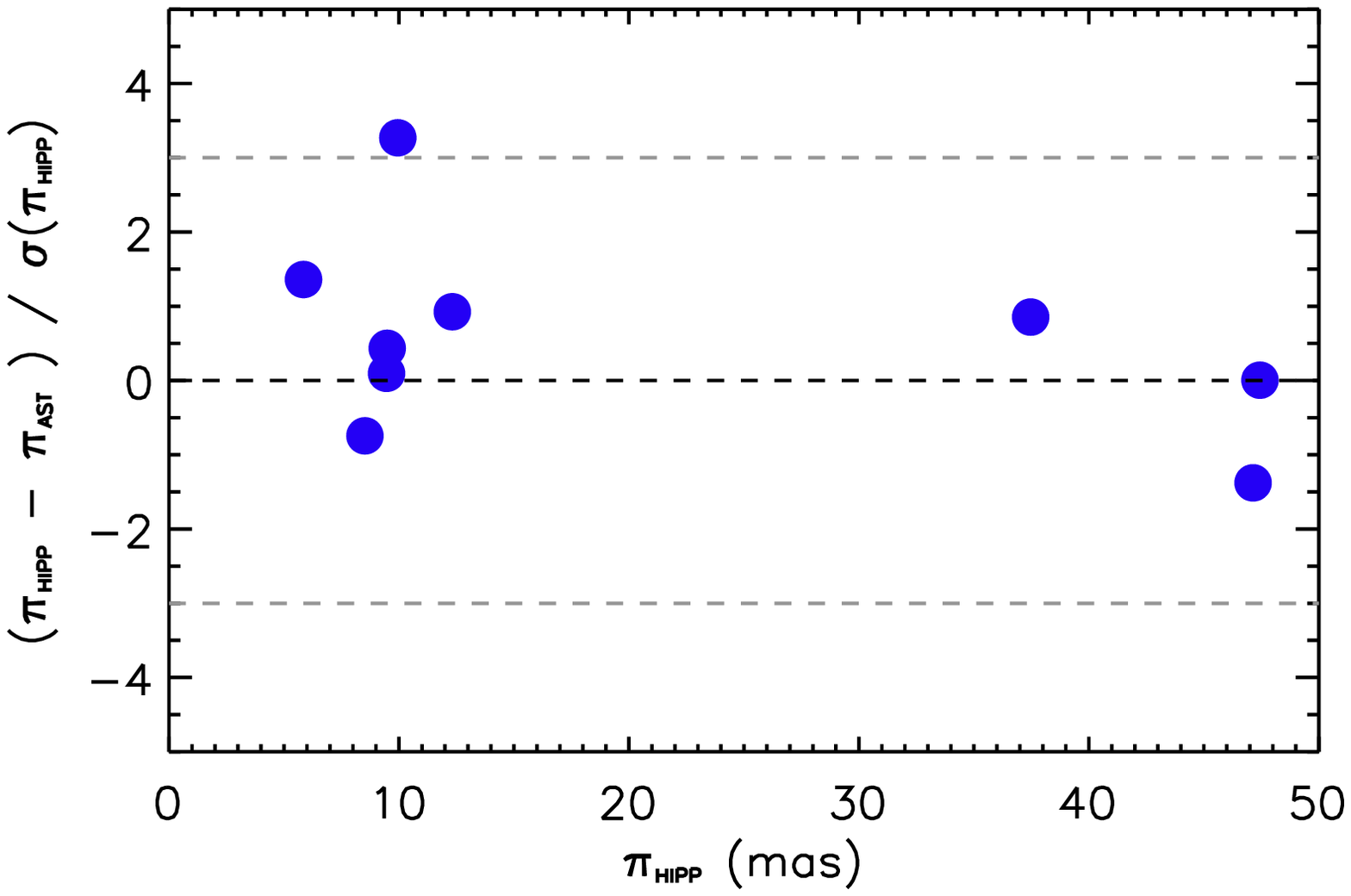}
    \caption{Comparison of measured radii (top), luminosities (middle),
      and parallaxes (lower) with those deduced from the asteroseismic
      parameters.  The interferometric radii are denoted by the red circles
    in the top panel, and the green triangle is the value from \citet{masana06}}
    \label{fig:lumrad}
\end{figure}

\section{Results\label{sec4}}
\begin{table*}
\begin{center}\caption{Reference models of the {\it Kepler} targets and the Sun \label{tab:referencemodels}}
\begin{tabular}{lccccccccccccccccccc}
\hline\hline
KIC ID& $R$ & $M$ & Age & $Z_i$ & $Y_i$  & $\alpha$ & $X_c/X_i$ & $a_0$ & $\chi^2_{N,r01}$ & $\chi^2_{N,r02}$ & $\chi^2_{N,\rm spec}$\\
    & (\rsol) & (\msol) & (Gyr) & & \\
\hline
Sun & 1.003 & 1.01 & 4.38 & 0.0210 & 0.281 & 2.24 & 0.50 & -2.54 & 1.03 & 0.78 & 0.71\\
1435467 & 1.704 & 1.41 & 1.87 & 0.0231 & 0.284 & 1.84 & 0.43 & -3.95 & 2.68 & 1.64 & 1.49\\
2837475 & 1.613 & 1.41 & 1.70 & 0.0168 & 0.247 & 1.70 & 0.53 & -4.48 & 1.29 & 2.07 & 0.32\\
3427720 & 1.125 & 1.13 & 2.17 & 0.0168 & 0.259 & 2.10 & 0.64 & -2.41 & 1.10 & 1.26 & 0.15\\
3656476 & 1.326 & 1.10 & 8.48 & 0.0231 & 0.248 & 2.30 & 0.00 & -2.22 & 2.35 & 0.68 & 1.57\\
3735871 & 1.089 & 1.08 & 1.57 & 0.0157 & 0.292 & 2.02 & 0.71 & -3.64 & 1.47 & 0.67 & 0.05\\
4914923 & 1.326 & 1.01 & 7.15 & 0.0121 & 0.260 & 1.68 & 0.02 & -4.51 & 0.56 & 1.50 & 3.35\\
5184732 & 1.365 & 1.27 & 4.70 & 0.0340 & 0.242 & 1.92 & 0.27 & -4.43 & 6.98 & 2.32 & 0.85\\
5950854 & 1.257 & 1.01 & 9.01 & 0.0147 & 0.249 & 2.16 & 0.00 & -1.27 & 0.60 & 4.61 & 1.30\\
6106415 & 1.213 & 1.06 & 4.43 & 0.0184 & 0.295 & 2.04 & 0.18 & -3.48 & 0.93 & 2.81 & 0.54\\
6116048 & 1.239 & 1.06 & 5.84 & 0.0114 & 0.242 & 2.16 & 0.11 & -3.27 & 3.27 & 2.48 & 0.44\\
6225718 & 1.194 & 1.06 & 2.30 & 0.0117 & 0.286 & 2.02 & 0.49 & -5.99 & 3.47 & 0.97 & 0.64\\
6603624 & 1.159 & 1.03 & 8.64 & 0.0455 & 0.313 & 2.12 & 0.01 & -2.34 & 3.42 & 135.14 & 5.90\\
6933899 & 1.535 & 1.03 & 6.58 & 0.0152 & 0.296 & 1.76 & 0.00 & -4.38 & 1.45 & 1.25 & 0.21\\
7103006 & 1.957 & 1.56 & 1.94 & 0.0224 & 0.239 & 1.66 & 0.36 & -7.28 & 1.15 & 0.69 & 1.33\\
7106245 & 1.120 & 0.97 & 6.05 & 0.0070 & 0.242 & 1.98 & 0.22 & -4.02 & 2.96 & 0.73 & 4.41\\
7206837 & 1.579 & 1.41 & 1.72 & 0.0255 & 0.249 & 1.52 & 0.60 & -4.61 & 1.48 & 1.43 & 1.52\\
7296438 & 1.371 & 1.10 & 5.93 & 0.0309 & 0.315 & 2.04 & 0.02 & -2.76 & 0.74 & 0.53 & 0.47\\
7510397 & 1.828 & 1.30 & 3.58 & 0.0129 & 0.248 & 1.84 & 0.08 & -2.37 & 0.75 & 2.23 & 0.55\\
7680114 & 1.395 & 1.07 & 7.04 & 0.0197 & 0.277 & 2.02 & 0.00 & -3.00 & 1.63 & 0.74 & 0.00\\
7771282 & 1.645 & 1.30 & 3.13 & 0.0168 & 0.257 & 1.78 & 0.19 & -4.03 & 2.10 & 0.75 & 0.33\\
7871531 & 0.859 & 0.80 & 9.32 & 0.0125 & 0.296 & 2.02 & 0.34 & -4.15 & 1.06 & 0.65 & 1.25\\
7940546 & 1.917 & 1.39 & 2.58 & 0.0152 & 0.259 & 1.74 & 0.07 & -6.26 & 2.47 & 0.82 & 1.45\\
7970740 & 0.779 & 0.78 & 10.59 & 0.0094 & 0.244 & 2.36 & 0.45 & -2.55 & 4.93 & 5.09 & 3.34\\
8006161 & 0.954 & 1.06 & 4.34 & 0.0485 & 0.288 & 2.66 & 0.61 & -0.63 & 2.33 & 1.21 & 1.26\\
8150065 & 1.394 & 1.20 & 3.33 & 0.0162 & 0.252 & 1.62 & 0.21 & -3.97 & 2.03 & 2.30 & 0.66\\
8179536 & 1.353 & 1.26 & 2.03 & 0.0157 & 0.249 & 1.88 & 0.50 & -3.89 & 1.51 & 0.62 & 0.01\\
8379927 & 1.105 & 1.08 & 1.65 & 0.0162 & 0.287 & 1.82 & 0.71 & -4.98 & 1.87 & 1.63 & 0.33\\
8394589 & 1.169 & 1.06 & 3.82 & 0.0094 & 0.247 & 1.98 & 0.37 & -3.14 & 0.71 & 0.70 & 0.01\\
8424992 & 1.056 & 0.94 & 9.62 & 0.0162 & 0.264 & 2.30 & 0.14 & -1.38 & 0.70 & 0.30 & 0.22\\
8694723 & 1.493 & 1.04 & 4.22 & 0.0085 & 0.309 & 2.36 & 0.00 & -2.23 & 0.70 & 1.46 & 3.18\\
8760414 & 1.028 & 0.82 & 12.09 & 0.0042 & 0.239 & 2.14 & 0.07 & -2.42 & 0.52 & 1.69 & 4.43\\
8938364 & 1.361 & 1.00 & 11.00 & 0.0217 & 0.272 & 2.14 & 0.00 & -2.09 & 1.44 & 3.52 & 3.26\\
9025370 & 1.000 & 0.97 & 5.50 & 0.0184 & 0.253 & 1.60 & 0.54 & -6.01 & 1.45 & 3.78 & 0.27\\
9098294 & 1.151 & 0.99 & 8.22 & 0.0129 & 0.245 & 2.14 & 0.11 & -3.13 & 1.93 & 0.96 & 0.23\\
9139151 & 1.167 & 1.20 & 1.84 & 0.0203 & 0.265 & 2.48 & 0.63 & -1.58 & 1.66 & 1.26 & 0.17\\
9139163 & 1.582 & 1.49 & 1.26 & 0.0330 & 0.245 & 1.64 & 0.71 & -9.60 & 0.95 & 1.89 & 4.25\\
9206432 & 1.499 & 1.37 & 1.32 & 0.0247 & 0.285 & 1.82 & 0.65 & -2.37 & 1.68 & 1.10 & 0.72\\
9353712 & 2.183 & 1.56 & 2.17 & 0.0203 & 0.249 & 1.76 & 0.08 & -1.89 & 2.57 & 0.73 & 1.16\\
9410862 & 1.159 & 0.99 & 6.15 & 0.0091 & 0.247 & 1.90 & 0.20 & -3.11 & 1.28 & 0.75 & 0.74\\
9414417 & 1.896 & 1.40 & 2.67 & 0.0147 & 0.244 & 1.70 & 0.11 & -5.41 & 1.01 & 0.78 & 0.39\\
9955598 & 0.876 & 0.87 & 6.38 & 0.0203 & 0.308 & 2.16 & 0.48 & -2.71 & 1.15 & 2.13 & 0.13\\
9965715 & 1.224 & 0.99 & 3.00 & 0.0080 & 0.310 & 1.58 & 0.33 & -5.57 & 0.78 & 0.65 & 1.76\\
10079226 & 1.135 & 1.09 & 2.35 & 0.0203 & 0.291 & 1.84 & 0.61 & -4.10 & 1.39 & 0.73 & 0.12\\
10454113 & 1.282 & 1.27 & 2.03 & 0.0217 & 0.244 & 2.02 & 0.58 & -0.79 & 2.07 & 4.38 & 1.79\\
10516096 & 1.407 & 1.08 & 6.44 & 0.0168 & 0.270 & 2.04 & 0.00 & -2.81 & 1.29 & 1.14 & 0.65\\
10644253 & 1.073 & 1.04 & 1.14 & 0.0162 & 0.319 & 1.78 & 0.78 & -4.91 & 0.78 & 0.62 & 0.31\\
10730618 & 1.729 & 1.33 & 2.55 & 0.0147 & 0.253 & 1.34 & 0.30 & -2.14 & 2.04 & 3.36 & 0.14\\
10963065 & 1.210 & 1.04 & 4.28 & 0.0114 & 0.277 & 2.04 & 0.22 & -3.53 & 1.41 & 0.98 & 0.00\\
11081729 & 1.393 & 1.25 & 1.88 & 0.0143 & 0.271 & 1.86 & 0.51 & -5.62 & 6.03 & 5.17 & 1.56\\
11253226 & 1.635 & 1.53 & 1.06 & 0.0224 & 0.248 & 1.90 & 0.69 & -4.76 & 2.76 & 1.83 & 2.00\\
11772920 & 0.839 & 0.81 & 11.11 & 0.0143 & 0.254 & 1.82 & 0.43 & -3.90 & 2.28 & 0.35 & 0.33\\
12009504 & 1.379 & 1.13 & 3.44 & 0.0157 & 0.294 & 1.96 & 0.26 & -4.67 & 0.81 & 0.88 & 0.10\\
12069127 & 2.262 & 1.58 & 1.89 & 0.0203 & 0.262 & 1.64 & 0.12 & -4.46 & 3.00 & 0.79 & 0.02\\
12069424 & 1.223 & 1.07 & 7.35 & 0.0179 & 0.241 & 2.12 & 0.09 & -4.41 & 3.78 & 1.02 & 1.39\\
12069449 & 1.105 & 1.01 & 6.88 & 0.0217 & 0.278 & 2.14 & 0.22 & -2.90 & 4.93 & 0.94 & 0.69\\
12258514 & 1.601 & 1.25 & 6.11 & 0.0247 & 0.229 & 1.64 & 0.00 & -4.04 & 2.45 & 0.92 & 9.89\\
12317678 & 1.749 & 1.27 & 2.18 & 0.0107 & 0.302 & 1.74 & 0.13 & -5.26 & 1.22 & 1.09 & 0.65\\
\hline\hline
\end{tabular}
\end{center}
Notes: The parameters are radius, mass, age, initial metallicity $Z_i$ and helium $Y_i$ mass fraction, mixing-length parameter $\alpha$, ratio of current central hydrogen to initial hydrogen mass fraction, $X_c/X_i$, the $a_0$ parameter in Eq.~\ref{eqn:kjeldsen}, and the normalized $\chi^2$ values for the $r_{01}$, $r_{02}$ and spectroscopic data.
\end{table*}

{The sample of stars analyzed in this work span  the 
main-sequence and early subgiant phase, as illustrated by their position in the
Hertzsprung-Russell diagram (Fig.~\ref{fig:hrdiag}).
They cover a range in mass of about 0.6 \msol, with about half of the sample
being within 10\% of the solar value. 
For a representative set of four stars, Fig.~\ref{fig:fitdata0} compares
the measured frequency separation ratios (crosses) with the 
corresponding values from the reference models (red filled dots).  
Here it can be seen that the agreement with the seismic observations is 
in general excellent, but some of the models do not necessarily
reproduce features of the observed data. One example is KIC~10454113,
which is shown in the lower right panel. It displays an oscillation as a function of
frequency that the models 
fail to reproduce.  These discrepancies are indeed noted in the normalized
\chisq\ value, $\chi^2_N = \chi^2/N = 3.2$, where $N$ is the number of frequency
ratios.
For KIC~8006161, shown in the top left panel, the fit is of higher quality
with $\chi^2_N = 1.8$.}
The parameters of the reference models that are used to compare with the 
observations are listed in Table~\ref{tab:referencemodels} along with
the individual $\chi^2_N$ values for $r_{01}$, $r_{02}$, and combined
\teff\ and \mh.

{For the Sun and each star in our sample,  we derived a best estimate and uncertainty 
for the stellar radius, mass, age, metallicity, luminosity, and surface gravity using the 
method described in Sec.~\ref{sec3}
 (see Table~\ref{tab:properties_derived}). 
Using the rotation periods given in Table~\ref{tab1} and the derived
radius, we also
computed their rotational velocities.}

{Since the AMP~1.3 method uses only one set of physics in the stellar modeling}, 
the derived uncertainties 
do not
include possible systematic errors arising from errors in the 
model physics, such as the equation of state, heavy element settling, and convective overshoot. 
However, the uncertainties include sources of errors arising
from free parameters that are often fixed in {the stellar codes used in other methods}, for example,
the mixing-length parameter $\alpha$, the initial chemical composition $(X_i, Y_i, Z_i),$ or a chemical enrichment law.
The {uncertainty on these} parameters contributes substantially to the error budget, and in some 
cases more so, for example, changing
the equation of state or the opacities.
The effect of such changes in the physics has been studied in detail
for HD\,52265 by \cite{lebreton2014}.  A similar detailed {analysis} for each
star in the sample we studied is beyond the scope of this paper.  
{We refer to \citet{silvaaguirre2016}, who also analyzed data from \citet{Lund2016} 
using seven distinct modeling methods and codes.} 

The accuracy, namely the bias and not {the} precision,  of our results
can be {ascertained} by an analysis of the solar observations.  
As {stated} above, we {derived a best-matched model with
values for the mass of} a 1 \msol\ model and a radius of 1 \rsol, and
an age that, within the derived uncertainty, matches the solar value.
A second accuracy test, at least for the age, 
can be established based on the independently derived ages for the 
binary system 16 Cyg A and B (also known as KIC~12069449 and KIC~12069424). 
{The ages that we derive agree to within 1$\sigma$.}



\renewcommand{\tabcolsep}{4.2pt}
\begin{table*}
\caption{Derived stellar properties of the {\it Kepler} targets and the 
Sun using VIRGO data \label{tab:properties_derived}}
\begin{tabular}{lcccccccccccccccccccccc}
\hline\hline
KIC ID& $R$ & $M$ & Age & $L$ & $T_{\rm eff}$ & $\log g$ & [M/H] & $\pi$ & $v$\\
& (\rsol) & (\msol) & (Gyr) & (\lsol) & (K) & (dex) & (dex) & (mas) & (km~s$^{-1}$)\\
\hline
Sun & 1.001 $\pm$ 0.005 & 1.001 $\pm$ 0.019 & 4.38 $\pm$ 0.22 & 0.97 $\pm$ 0.03 & 5732 $\pm$ 43 & 4.438 $\pm$ 0.003 & 0.07 $\pm$ 0.04 &   &  \\
1435467 & 1.728 $\pm$ 0.027 & 1.466 $\pm$ 0.060 & 1.97 $\pm$ 0.17 & 4.29 $\pm$ 0.25 & 6299 $\pm$ 75 & 4.128 $\pm$ 0.004 & 0.09 $\pm$ 0.09 & 6.99 $\pm$ 0.24 & 13.09 $\pm$ 1.76\\
2837475 & 1.629 $\pm$ 0.027 & 1.460 $\pm$ 0.062 & 1.49 $\pm$ 0.22 & 4.54 $\pm$ 0.26 & 6600 $\pm$ 71 & 4.174 $\pm$ 0.007 & 0.05 $\pm$ 0.07 & 8.18 $\pm$ 0.29 & 22.40 $\pm$ 2.22\\
3427720 & 1.089 $\pm$ 0.009 & 1.034 $\pm$ 0.015 & 2.37 $\pm$ 0.23 & 1.37 $\pm$ 0.08 & 5989 $\pm$ 71 & 4.378 $\pm$ 0.003 & -0.05 $\pm$ 0.09 & 11.04 $\pm$ 0.40 & 3.95 $\pm$ 0.61\\
3656476 & 1.322 $\pm$ 0.007 & 1.101 $\pm$ 0.025 & 8.88 $\pm$ 0.41 & 1.63 $\pm$ 0.06 & 5690 $\pm$ 53 & 4.235 $\pm$ 0.004 & 0.17 $\pm$ 0.07 & 8.49 $\pm$ 0.30 & 2.11 $\pm$ 0.24\\
3735871 & 1.080 $\pm$ 0.012 & 1.068 $\pm$ 0.035 & 1.55 $\pm$ 0.18 & 1.45 $\pm$ 0.09 & 6092 $\pm$ 75 & 4.395 $\pm$ 0.005 & -0.05 $\pm$ 0.04 & 8.05 $\pm$ 0.31 & 4.74 $\pm$ 0.51\\
4914923 & 1.339 $\pm$ 0.015 & 1.039 $\pm$ 0.028 & 7.04 $\pm$ 0.50 & 1.79 $\pm$ 0.12 & 5769 $\pm$ 86 & 4.198 $\pm$ 0.004 & -0.06 $\pm$ 0.09 & 8.64 $\pm$ 0.35 & 3.31 $\pm$ 0.46\\
5184732 & 1.354 $\pm$ 0.028 & 1.247 $\pm$ 0.071 & 4.32 $\pm$ 0.85 & 1.79 $\pm$ 0.15 & 5752 $\pm$ 101 & 4.268 $\pm$ 0.009 & 0.31 $\pm$ 0.06 & 14.53 $\pm$ 0.67 & 3.46 $\pm$ 0.43\\
5950854 & 1.254 $\pm$ 0.012 & 1.005 $\pm$ 0.035 & 9.25 $\pm$ 0.68 & 1.58 $\pm$ 0.11 & 5780 $\pm$ 74 & 4.245 $\pm$ 0.006 & -0.11 $\pm$ 0.06 & 4.41 $\pm$ 0.18 &  \\
6106415 & 1.205 $\pm$ 0.009 & 1.039 $\pm$ 0.021 & 4.55 $\pm$ 0.28 & 1.61 $\pm$ 0.09 & 5927 $\pm$ 63 & 4.294 $\pm$ 0.003 & -0.00 $\pm$ 0.04 & 25.35 $\pm$ 0.87 &  \\
6116048 & 1.233 $\pm$ 0.011 & 1.048 $\pm$ 0.028 & 6.08 $\pm$ 0.40 & 1.77 $\pm$ 0.13 & 5993 $\pm$ 73 & 4.276 $\pm$ 0.003 & -0.20 $\pm$ 0.08 & 13.31 $\pm$ 0.57 & 3.61 $\pm$ 0.41\\
6225718 & 1.234 $\pm$ 0.018 & 1.169 $\pm$ 0.039 & 2.23 $\pm$ 0.20 & 2.08 $\pm$ 0.11 & 6252 $\pm$ 63 & 4.321 $\pm$ 0.005 & -0.09 $\pm$ 0.06 & 19.32 $\pm$ 0.60 &  \\
6603624 & 1.164 $\pm$ 0.024 & 1.058 $\pm$ 0.075 & 8.66 $\pm$ 0.68 & 1.23 $\pm$ 0.11 & 5644 $\pm$ 91 & 4.326 $\pm$ 0.008 & 0.24 $\pm$ 0.05 & 11.89 $\pm$ 0.59 &  \\
6933899 & 1.597 $\pm$ 0.008 & 1.155 $\pm$ 0.011 & 7.22 $\pm$ 0.53 & 2.63 $\pm$ 0.06 & 5815 $\pm$ 47 & 4.093 $\pm$ 0.002 & 0.11 $\pm$ 0.03 & 6.48 $\pm$ 0.15 &  \\
7103006 & 1.958 $\pm$ 0.025 & 1.568 $\pm$ 0.051 & 1.69 $\pm$ 0.12 & 5.58 $\pm$ 0.36 & 6332 $\pm$ 89 & 4.048 $\pm$ 0.006 & 0.09 $\pm$ 0.10 & 6.19 $\pm$ 0.23 & 21.44 $\pm$ 2.25\\
7106245 & 1.125 $\pm$ 0.009 & 0.989 $\pm$ 0.023 & 6.05 $\pm$ 0.39 & 1.56 $\pm$ 0.09 & 6078 $\pm$ 74 & 4.327 $\pm$ 0.003 & -0.44 $\pm$ 0.11 & 4.98 $\pm$ 0.20 &  \\
7206837 & 1.556 $\pm$ 0.018 & 1.377 $\pm$ 0.039 & 1.55 $\pm$ 0.50 & 3.37 $\pm$ 0.15 & 6269 $\pm$ 87 & 4.191 $\pm$ 0.008 & 0.07 $\pm$ 0.15 & 5.28 $\pm$ 0.15 & 19.49 $\pm$ 1.37\\
7296438 & 1.370 $\pm$ 0.009 & 1.099 $\pm$ 0.022 & 6.37 $\pm$ 0.60 & 1.85 $\pm$ 0.08 & 5754 $\pm$ 55 & 4.205 $\pm$ 0.003 & 0.21 $\pm$ 0.07 & 6.09 $\pm$ 0.18 & 2.76 $\pm$ 0.30\\
7510397 & 1.823 $\pm$ 0.018 & 1.309 $\pm$ 0.037 & 3.51 $\pm$ 0.24 & 4.19 $\pm$ 0.20 & 6119 $\pm$ 69 & 4.031 $\pm$ 0.004 & -0.14 $\pm$ 0.06 & 11.75 $\pm$ 0.36 &  \\
7680114 & 1.402 $\pm$ 0.014 & 1.092 $\pm$ 0.030 & 6.89 $\pm$ 0.46 & 2.07 $\pm$ 0.09 & 5833 $\pm$ 47 & 4.181 $\pm$ 0.004 & 0.08 $\pm$ 0.07 & 5.73 $\pm$ 0.17 & 2.70 $\pm$ 0.19\\
7771282 & 1.629 $\pm$ 0.016 & 1.268 $\pm$ 0.040 & 2.78 $\pm$ 0.47 & 3.61 $\pm$ 0.18 & 6223 $\pm$ 73 & 4.118 $\pm$ 0.004 & -0.03 $\pm$ 0.07 & 3.24 $\pm$ 0.10 & 6.94 $\pm$ 0.54\\
7871531 & 0.871 $\pm$ 0.008 & 0.834 $\pm$ 0.021 & 8.84 $\pm$ 0.46 & 0.60 $\pm$ 0.05 & 5482 $\pm$ 69 & 4.478 $\pm$ 0.006 & -0.16 $\pm$ 0.04 & 16.81 $\pm$ 0.81 & 1.31 $\pm$ 0.10\\
7940546 & 1.974 $\pm$ 0.045 & 1.511 $\pm$ 0.087 & 2.42 $\pm$ 0.17 & 5.69 $\pm$ 0.35 & 6330 $\pm$ 43 & 4.023 $\pm$ 0.005 & 0.00 $\pm$ 0.06 & 12.16 $\pm$ 0.44 & 8.79 $\pm$ 0.76\\
7970740 & 0.776 $\pm$ 0.007 & 0.768 $\pm$ 0.019 & 10.53 $\pm$ 0.43 & 0.42 $\pm$ 0.04 & 5282 $\pm$ 93 & 4.546 $\pm$ 0.003 & -0.37 $\pm$ 0.09 & 36.83 $\pm$ 1.71 & 2.19 $\pm$ 0.38\\
8006161 & 0.930 $\pm$ 0.009 & 1.000 $\pm$ 0.030 & 4.57 $\pm$ 0.36 & 0.64 $\pm$ 0.03 & 5351 $\pm$ 49 & 4.498 $\pm$ 0.003 & 0.41 $\pm$ 0.04 & 37.89 $\pm$ 1.18 & 1.58 $\pm$ 0.16\\
8150065 & 1.402 $\pm$ 0.018 & 1.222 $\pm$ 0.040 & 3.15 $\pm$ 0.49 & 2.52 $\pm$ 0.19 & 6138 $\pm$ 105 & 4.230 $\pm$ 0.005 & -0.04 $\pm$ 0.15 & 3.94 $\pm$ 0.18 &  \\
8179536 & 1.350 $\pm$ 0.013 & 1.249 $\pm$ 0.031 & 1.88 $\pm$ 0.25 & 2.63 $\pm$ 0.11 & 6318 $\pm$ 59 & 4.274 $\pm$ 0.005 & -0.04 $\pm$ 0.07 & 6.91 $\pm$ 0.20 & 2.78 $\pm$ 0.18\\
8379927 & 1.102 $\pm$ 0.012 & 1.073 $\pm$ 0.033 & 1.64 $\pm$ 0.12 & 1.39 $\pm$ 0.10 & 5971 $\pm$ 91 & 4.382 $\pm$ 0.005 & -0.04 $\pm$ 0.05 & 30.15 $\pm$ 1.40 & 3.28 $\pm$ 0.26\\
8394589 & 1.155 $\pm$ 0.009 & 1.024 $\pm$ 0.030 & 3.82 $\pm$ 0.25 & 1.68 $\pm$ 0.09 & 6103 $\pm$ 61 & 4.324 $\pm$ 0.003 & -0.28 $\pm$ 0.07 & 8.47 $\pm$ 0.28 &  \\
8424992 & 1.048 $\pm$ 0.005 & 0.930 $\pm$ 0.016 & 9.79 $\pm$ 0.76 & 0.99 $\pm$ 0.04 & 5634 $\pm$ 57 & 4.362 $\pm$ 0.002 & -0.12 $\pm$ 0.06 & 7.52 $\pm$ 0.23 &  \\
8694723 & 1.463 $\pm$ 0.023 & 1.004 $\pm$ 0.036 & 4.85 $\pm$ 0.22 & 3.15 $\pm$ 0.18 & 6347 $\pm$ 67 & 4.107 $\pm$ 0.004 & -0.38 $\pm$ 0.08 & 8.18 $\pm$ 0.28 &  \\
8760414 & 1.027 $\pm$ 0.004 & 0.814 $\pm$ 0.011 & 11.88 $\pm$ 0.34 & 1.15 $\pm$ 0.06 & 5915 $\pm$ 54 & 4.329 $\pm$ 0.002 & -0.66 $\pm$ 0.07 & 9.83 $\pm$ 0.32 &  \\
8938364 & 1.362 $\pm$ 0.007 & 1.015 $\pm$ 0.023 & 10.85 $\pm$ 1.22 & 1.65 $\pm$ 0.15 & 5604 $\pm$ 115 & 4.174 $\pm$ 0.004 & 0.06 $\pm$ 0.06 & 6.27 $\pm$ 0.31 &  \\
9025370 & 0.997 $\pm$ 0.017 & 0.969 $\pm$ 0.036 & 5.53 $\pm$ 0.43 & 0.71 $\pm$ 0.11 & 5296 $\pm$ 157 & 4.424 $\pm$ 0.006 & 0.01 $\pm$ 0.09 & 15.66 $\pm$ 1.44 &  \\
9098294 & 1.150 $\pm$ 0.003 & 0.979 $\pm$ 0.017 & 8.23 $\pm$ 0.53 & 1.34 $\pm$ 0.05 & 5795 $\pm$ 53 & 4.312 $\pm$ 0.002 & -0.17 $\pm$ 0.07 & 8.30 $\pm$ 0.23 & 2.94 $\pm$ 0.20\\
9139151 & 1.137 $\pm$ 0.027 & 1.129 $\pm$ 0.091 & 1.94 $\pm$ 0.31 & 1.81 $\pm$ 0.11 & 6270 $\pm$ 63 & 4.375 $\pm$ 0.008 & 0.05 $\pm$ 0.10 & 9.57 $\pm$ 0.34 & 5.25 $\pm$ 1.07\\
9139163 & 1.569 $\pm$ 0.027 & 1.480 $\pm$ 0.085 & 1.23 $\pm$ 0.15 & 3.51 $\pm$ 0.24 & 6318 $\pm$ 105 & 4.213 $\pm$ 0.004 & 0.11 $\pm$ 0.00 & 9.85 $\pm$ 0.39 &  \\
9206432 & 1.460 $\pm$ 0.015 & 1.301 $\pm$ 0.048 & 1.48 $\pm$ 0.31 & 3.47 $\pm$ 0.18 & 6508 $\pm$ 75 & 4.219 $\pm$ 0.009 & 0.06 $\pm$ 0.07 & 7.03 $\pm$ 0.26 & 8.39 $\pm$ 1.01\\
9353712 & 2.240 $\pm$ 0.061 & 1.681 $\pm$ 0.125 & 1.91 $\pm$ 0.14 & 7.27 $\pm$ 1.02 & 6343 $\pm$ 119 & 3.965 $\pm$ 0.008 & 0.12 $\pm$ 0.08 & 2.21 $\pm$ 0.16 & 10.03 $\pm$ 1.03\\
9410862 & 1.149 $\pm$ 0.009 & 0.969 $\pm$ 0.017 & 5.78 $\pm$ 0.82 & 1.56 $\pm$ 0.08 & 6017 $\pm$ 69 & 4.304 $\pm$ 0.003 & -0.34 $\pm$ 0.08 & 5.05 $\pm$ 0.16 & 2.55 $\pm$ 0.27\\
9414417 & 1.891 $\pm$ 0.015 & 1.401 $\pm$ 0.028 & 2.53 $\pm$ 0.17 & 4.98 $\pm$ 0.22 & 6260 $\pm$ 67 & 4.028 $\pm$ 0.004 & -0.07 $\pm$ 0.12 & 4.65 $\pm$ 0.13 & 8.96 $\pm$ 0.56\\
9955598 & 0.881 $\pm$ 0.008 & 0.885 $\pm$ 0.023 & 6.47 $\pm$ 0.45 & 0.58 $\pm$ 0.03 & 5400 $\pm$ 57 & 4.494 $\pm$ 0.003 & 0.06 $\pm$ 0.04 & 14.98 $\pm$ 0.53 & 1.30 $\pm$ 0.22\\
9965715 & 1.234 $\pm$ 0.015 & 1.005 $\pm$ 0.033 & 3.29 $\pm$ 0.33 & 1.85 $\pm$ 0.15 & 6058 $\pm$ 113 & 4.258 $\pm$ 0.004 & -0.27 $\pm$ 0.11 & 8.81 $\pm$ 0.51 &  \\
10079226 & 1.129 $\pm$ 0.016 & 1.082 $\pm$ 0.048 & 2.75 $\pm$ 0.42 & 1.41 $\pm$ 0.10 & 5915 $\pm$ 89 & 4.364 $\pm$ 0.005 & 0.07 $\pm$ 0.06 & 7.05 $\pm$ 0.29 & 3.86 $\pm$ 0.33\\
10454113 & 1.272 $\pm$ 0.006 & 1.260 $\pm$ 0.016 & 2.06 $\pm$ 0.16 & 2.07 $\pm$ 0.08 & 6134 $\pm$ 61 & 4.325 $\pm$ 0.003 & 0.04 $\pm$ 0.04 & 11.94 $\pm$ 0.63 & 4.41 $\pm$ 0.33\\
10516096 & 1.398 $\pm$ 0.008 & 1.065 $\pm$ 0.012 & 6.59 $\pm$ 0.37 & 2.11 $\pm$ 0.08 & 5872 $\pm$ 43 & 4.173 $\pm$ 0.003 & -0.06 $\pm$ 0.06 & 7.53 $\pm$ 0.21 &  \\
10644253 & 1.090 $\pm$ 0.027 & 1.091 $\pm$ 0.097 & 0.94 $\pm$ 0.26 & 1.45 $\pm$ 0.09 & 6033 $\pm$ 67 & 4.399 $\pm$ 0.007 & 0.01 $\pm$ 0.10 & 10.45 $\pm$ 0.39 & 5.05 $\pm$ 0.42\\
10730618 & 1.763 $\pm$ 0.040 & 1.411 $\pm$ 0.097 & 1.81 $\pm$ 0.41 & 4.04 $\pm$ 0.56 & 6156 $\pm$ 181 & 4.095 $\pm$ 0.011 & 0.05 $\pm$ 0.18 & 3.35 $\pm$ 0.27 &  \\
10963065 & 1.204 $\pm$ 0.007 & 1.023 $\pm$ 0.024 & 4.33 $\pm$ 0.30 & 1.80 $\pm$ 0.08 & 6097 $\pm$ 53 & 4.288 $\pm$ 0.003 & -0.24 $\pm$ 0.06 & 11.46 $\pm$ 0.34 & 4.84 $\pm$ 0.65\\
11081729 & 1.423 $\pm$ 0.009 & 1.257 $\pm$ 0.045 & 2.22 $\pm$ 0.10 & 3.29 $\pm$ 0.07 & 6474 $\pm$ 43 & 4.215 $\pm$ 0.026 & 0.07 $\pm$ 0.03 & 7.48 $\pm$ 0.17 & 26.28 $\pm$ 2.98\\
11253226 & 1.606 $\pm$ 0.015 & 1.486 $\pm$ 0.030 & 0.97 $\pm$ 0.21 & 4.80 $\pm$ 0.20 & 6696 $\pm$ 79 & 4.197 $\pm$ 0.007 & 0.10 $\pm$ 0.05 & 8.07 $\pm$ 0.23 & 22.32 $\pm$ 2.28\\
11772920 & 0.845 $\pm$ 0.009 & 0.830 $\pm$ 0.028 & 10.79 $\pm$ 0.96 & 0.42 $\pm$ 0.06 & 5084 $\pm$ 159 & 4.502 $\pm$ 0.004 & -0.06 $\pm$ 0.09 & 14.82 $\pm$ 1.24 &  \\
12009504 & 1.382 $\pm$ 0.022 & 1.137 $\pm$ 0.063 & 3.44 $\pm$ 0.44 & 2.46 $\pm$ 0.25 & 6140 $\pm$ 133 & 4.213 $\pm$ 0.006 & -0.04 $\pm$ 0.05 & 7.51 $\pm$ 0.42 & 7.44 $\pm$ 0.55\\
12069127 & 2.283 $\pm$ 0.033 & 1.621 $\pm$ 0.084 & 1.79 $\pm$ 0.14 & 7.26 $\pm$ 0.42 & 6267 $\pm$ 79 & 3.926 $\pm$ 0.010 & 0.15 $\pm$ 0.08 & 2.35 $\pm$ 0.08 & 125.54 $\pm$ 7.07\\
12069424 & 1.223 $\pm$ 0.005 & 1.072 $\pm$ 0.013 & 7.36 $\pm$ 0.31 & 1.52 $\pm$ 0.05 & 5785 $\pm$ 39 & 4.294 $\pm$ 0.001 & -0.04 $\pm$ 0.05 & 47.44 $\pm$ 1.00 & 2.60 $\pm$ 0.20\\
12069449 & 1.113 $\pm$ 0.016 & 1.038 $\pm$ 0.047 & 7.05 $\pm$ 0.63 & 1.21 $\pm$ 0.11 & 5732 $\pm$ 83 & 4.361 $\pm$ 0.007 & 0.15 $\pm$ 0.08 & 46.77 $\pm$ 2.10 & 2.43 $\pm$ 0.63\\
12258514 & 1.593 $\pm$ 0.016 & 1.251 $\pm$ 0.016 & 5.50 $\pm$ 0.40 & 2.63 $\pm$ 0.12 & 5808 $\pm$ 61 & 4.129 $\pm$ 0.002 & 0.10 $\pm$ 0.09 & 12.79 $\pm$ 0.40 & 5.37 $\pm$ 0.66\\
12317678 & 1.788 $\pm$ 0.014 & 1.373 $\pm$ 0.030 & 2.30 $\pm$ 0.20 & 5.49 $\pm$ 0.28 & 6587 $\pm$ 97 & 4.064 $\pm$ 0.005 & -0.26 $\pm$ 0.09 & 6.89 $\pm$ 0.23 &  \\
\hline\hline
\end{tabular}
Notes: The mean model parameters are radius, mass, age, luminosity, effective temperature, surface gravity, metallicity, parallax, and rotational velocity.  
The latter two are derived using data from this table and Table~\ref{tab1}.
\end{table*}


\subsection{Accuracy of radii and luminosities\label{sec:comparison}}

To test the accuracy of the derived radii and luminosities, we have 
compiled measured values of these properties for nine stars (Table~\ref{tab:lumradobs}). {These stars have reliable Hipparcos parallaxes and are not members of 
close binary systems.}
Only three of the radii of the subsample of stars have been
measured interferometrically \citep{huber12,white13}. {The angular diameters from \citet{masana06} and \citet{huber14} were derived from broadband photometry and from literature atmospheric properties and stellar evolution models, respectively.}
\citet{met1216cyg} and \citet{Metcalfe2014} derived the luminosities using  
{extinction 
estimates from \citet{ammons2006} and the bolometric corrections from  \citeauthor[]{flower1996} (\citeyear{flower1996}, see \citealt{torres2010}).}

\begin{table}[]
    \begin{center}
    \caption{Luminosities, radii, and parallaxes from independent sources}
    \label{tab:lumradobs}
    \begin{tabular}{lcccccc}
    \hline\hline
        KIC ID & $L$  & $R$ & $\pi$ \\
        & (\lsol) & (\rsol) & (mas) \\
            \hline
8006161 & 0.61 $\pm$ 0.02 & 0.950$^1$ $\pm$ 0.020 & 37.47 $\pm$ 0.49 \\
9139151 & 1.63 $\pm$ 0.40 & 1.160$^3$ $\pm$ 0.020 & 9.46 $\pm$ 1.15 \\
9139163 & 3.88 $\pm$ 0.69 & 1.570$^3$ $\pm$ 0.030 & 9.49 $\pm$ 0.83 \\
9206432 & 4.95 $\pm$ 1.48 & 1.520$^3$ $\pm$ 0.030 & 5.85 $\pm$ 0.87 \\
10454113 & 2.60 $\pm$ 0.36 & 1.240$^3$ $\pm$ 0.020 & 9.95 $\pm$ 0.67 \\
11253226 & 4.22 $\pm$ 0.61 & 1.576$^4$ $\pm$ 0.143 & 8.52 $\pm$ 0.60 \\
12069424 & 1.56 $\pm$ 0.05 & 1.220$^1$ $\pm$ 0.020 & 47.44 $\pm$ 0.27 \\
12069449 & 1.27 $\pm$ 0.04 & 1.120$^1$ $\pm$ 0.020 & 47.14 $\pm$ 0.27 \\
12258514 & 2.84 $\pm$ 0.25 & 1.590$^3$ $\pm$ 0.040 & 12.32 $\pm$ 0.51 \\
         \hline\hline
    \end{tabular}
    \end{center}
The luminosities are from \citet{met1216cyg,Metcalfe2014}.  The references to 
the radii are 
$^1$\citet{huber12} $^2$\citet{white13} $^3$\citet{huber14} $^4$\citet{masana06}.
The parallaxes are from \citet{hipparcos07}.
\end{table}

A comparison of these independent measures of stellar radii and luminosities with
those derived using our asteroseismic methodology is shown in the top
two panels of Fig.~\ref{fig:lumrad}.
This comparison, using measurement differences relative to their
uncertainty as listed in the literature, shows no
systematic biases or trends for this subsample of nine stars.
The mean relative difference is --0.40 with a root mean
square (rms) around the 
mean of 0.59 
for the interferometrically measured radii (references 1 and 2, red filled circles)
and --0.28 $\pm$ 1.03 for the radii derived using photometry 
and isochrones (references 3 and 4).  
For the luminosity the mean relative difference is --0.35 with an rms around
the mean of 1.1.

\subsection{Asteroseismic parallaxes \label{sec:asteroseismicdistances}}
We used the luminosity $L$ that was derived from the asteroseismic analysis to compute
the stellar distance as a parallax.
Using the modeled surface gravity and the observed
\teff\ and \feh,\ we derived the amount of interstellar absorption between the top
of the Earth's atmosphere and the star, $A_{Ks}$, using 
the isochrone method described in \citet{schultheis2014}.  
Here, the subscript $Ks$ refers to the 2MASS $K_s$ filter \citep{2MASS}.
With the same observed \teff\ , we computed the corresponding 
bolometric correction $BC_{Ks}$ for this band,
using $BC_{Ks} = 4.51465 -0.000524461 T_{\rm eff}$ \citep{marigo2008} 
where the solar bolometric magnitude is 4.72 mag.
The $K_s$-band magnitude and $A_{Ks}$ are listed in Table~\ref{tab1}.
The distance, $d$, or parallax, $\pi$, 
is then computed directly from $L$, $K_s$, $BC_{Ks}$ , and $A_{Ks}$.


{The parallaxes and uncertainties of the stars in our sample
are listed in Table~\ref{tab:properties_derived}.
They were derived using Monte Carlo simulations, described as follows.
We perturbed each of the input data measures $L$, $A_{Ks}$, $K_s$, and $BC_K$,
using noise sampled from a Gaussian distribution with zero mean
and standard deviation equivalent to their errors to calculate 
a parallax.   
By repeating the perturbations 10,000 times, we obtained
a distribution of parallaxes, which is modeled by
a Gaussian function.   
The mean and standard deviation are adopted as the parallax
value and its uncertainty.
In most cases, the derived parallax error is dominated by the 
luminosity error.}

A comparison between the derived parallaxes and existing literature 
values (\citealt{hipparcos07}, Table~\ref{tab:lumradobs})
again validates our results, as shown in {the lower panel of} 
Fig.~\ref{fig:lumrad}, where no
significant trend can be seen.
In particular, we note that for the binary KIC~12069424 and KIC~12069449
(16 Cyg A\&B), we obtain almost identical parallaxes of 
{47.4 mas and 46.8 mas,} equivalent to 
a difference of 0.3 pc at a distance of 21.2 pc.  
{This result provides further evidence 
of the accuracy of our derived properties.}

\subsection{Trends in stellar properties \label{sec:stellartrends}}

Performing a homogenous analysis on a {relatively large sample
allows us to check for trends in some stellar parameters
and compare them to trends derived or established by other methods.
We performed this check for two parameters: the mixing-length
parameter and the stellar age.}

\subsubsection{Mixing-length parameter versus \teff\ and \logg}
The mixing-length parameter $\alpha$ is
usually calibrated for a solar model and then applied to all 
models for a set range of masses and metallicities.  
However, several authors have shown that this approach
is not correct, for instance, \citet{yildiz2006,bonaca2012,creevey2012A}. 
The values of $\alpha$ resulting from a GA {analysis} offer
an optimal approach to 
effectively test and subsequently constrain this parameter, since {by design}
the GA only restricts $\alpha$ to be between 1.0 and 3.0, {a range large enough
to encompass all plausible values}.

\begin{figure}
    \includegraphics[width=0.48\textwidth]{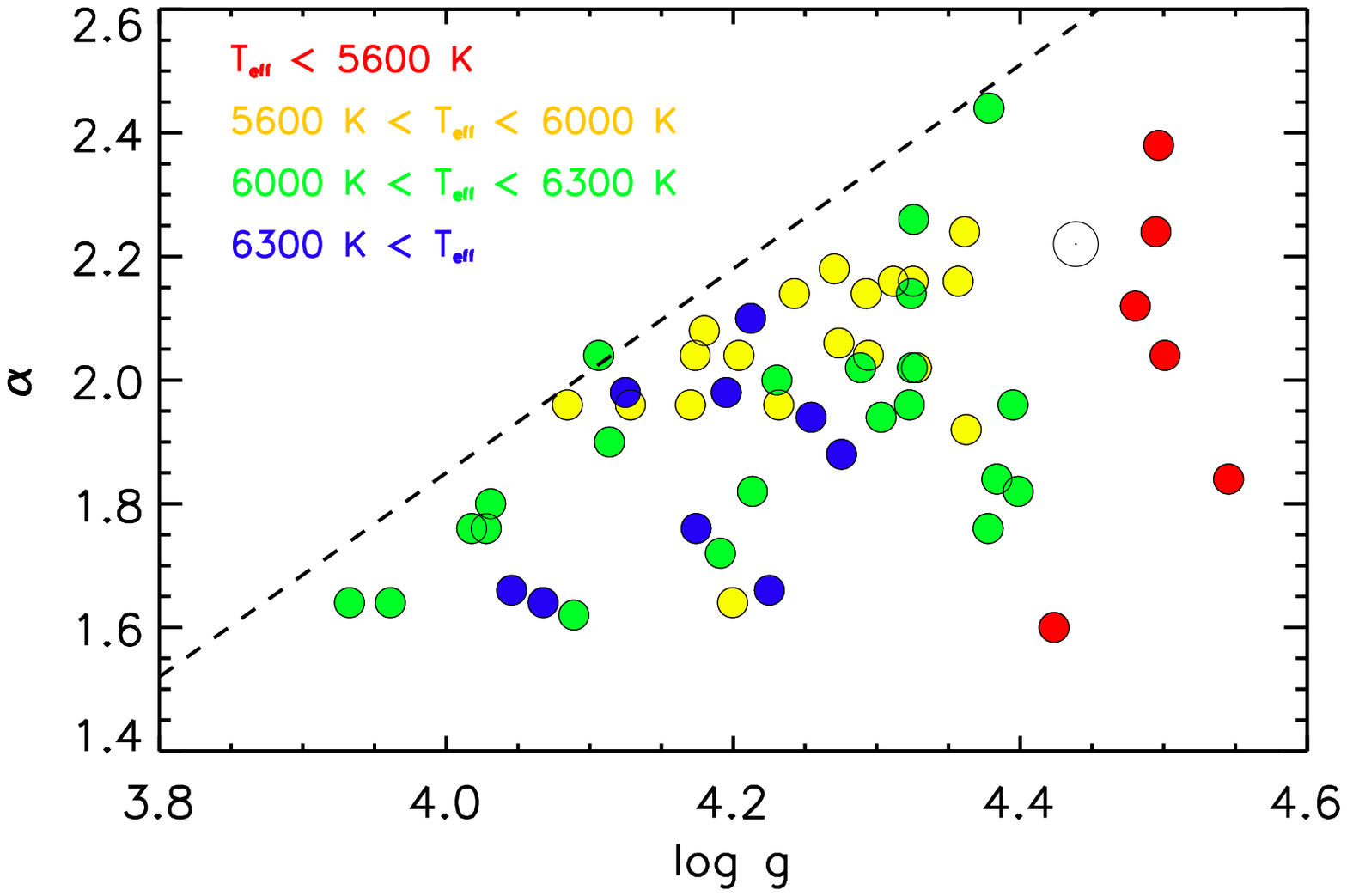}
    \includegraphics[width=0.48\textwidth]{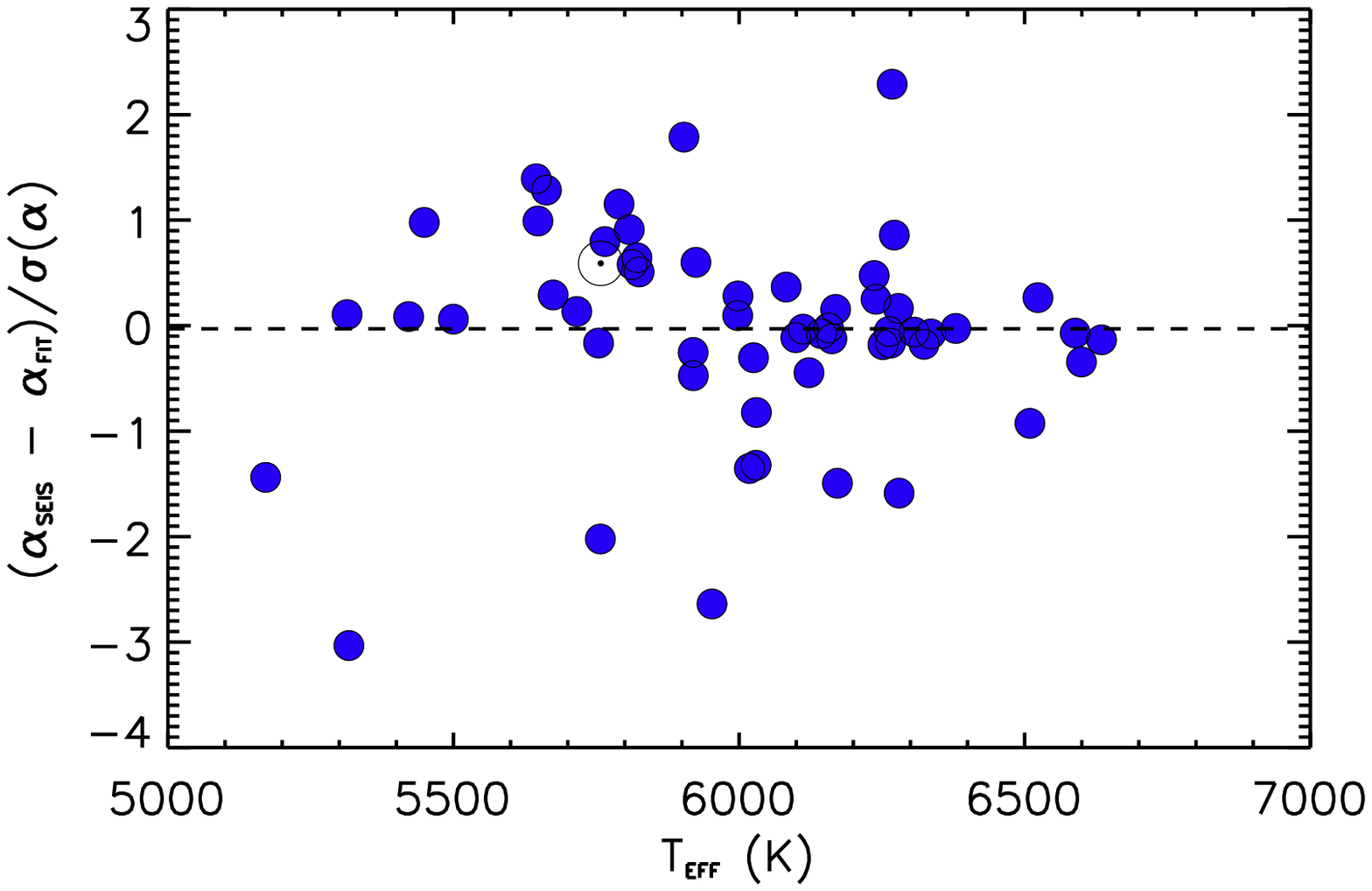}
    \caption{{\sl Top:} Distribution of the \logg\ and $\alpha$ for the full sample.
        The color coding shows in 
        red \teff\ $<$ 5600 K, 
        in yellow 5600 K $<$ \teff\ $<$ 6000 K,
        in green 6000 K $<$ \teff\ $<$ 6300 K,
        and blue \teff\ $>$ 6300 K.
        {\sl Bottom:} Residuals of the regression analysis scaled by the uncertainties
        in $\alpha$ as a function of \teff.
    \label{fig:poster_teffalpha}}
\end{figure}

The color-coded distribution of $\alpha$ 
with \logg\ and \teff\
{is shown}
in the top panel of Fig.~\ref{fig:poster_teffalpha}, using the
results derived from our sample of 57 stars and the Sun.
It is evident from this figure that for a given value of \logg, the 
value of $\alpha$ has an upper limit.
This upper limit can be represented by the equation 
$\alpha < 1.65 \log g - 4.75$, 
and this
is denoted by the dashed line in the figure.  
A regression analysis considering the model values of \logg, $\log$\teff\ and
[M/H]  yields 
\begin{eqnarray}
\indent \alpha &=& 5.972778 + 0.636997 \log g \nonumber \\
  && - 1.799968 \log T_{\rm eff} + 0.040094 [{\rm M}/{\rm H}], \nonumber \label{eqn:alpha_regression}
\\
\end{eqnarray}
with a mean and rms of the residual to the fit of --0.01 $\pm$ 0.15 
for the 58 stars.  
The residuals of this fit scaled by the uncertainties in $\alpha$
are shown in the lower panel of 
Fig.~\ref{fig:poster_teffalpha} as a function of \teff.  
No trend with this parameter can be seen.
This equation yields a value of $\alpha =  2.03$ for the known solar properties, within 1$\sigma$ of 
its mean value (2.12).

{These results agree in part with those derived by \citet{magic2015}, 
who used a full 3D radiative hydrodynamic simulation for modeling convective envelopes.}
These authors found that $\alpha$ increases with \logg\ and decreases with
\teff, {which is qualitatively} in agreement with our results.  
{The size of the variation that they inferred}, however, is smaller than the 
values we find.  
In our sample,  $\alpha$ varies between 1.7 and 2.4, 
while for the same range in \logg, \teff, \citeauthor{magic2015} see variations
in $\alpha$ from 1.9 to 2.3.
We note that the range of metallicity 
{in our sample is much smaller than the range in their work.
This could be the reason of the weak and opposite dependence on $\alpha$ that we find}.

\subsubsection{Age and $\langle r_{02} \rangle$}

\begin{figure}
    \includegraphics[width=0.48\textwidth]{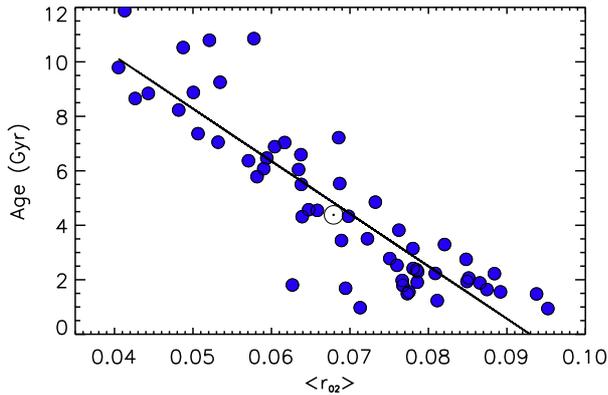}
    \caption{Age determination as a function of the mean value of $r02$. 
    \label{fig:res_Age_r02}}
\end{figure}

The $r_{02}$ frequency ratios {contain what is known as 
{\it \textup{small frequency separations,}}  
and} these are effective at probing the gradients near the core of the star \citep{Roxburgh2003}.
As the core is most sensitive to nuclear processing, $r_{02}$ are a diagnostic
of the evolutionary state of the star.  
Using theoretical models, \citet{lebMon2009} showed a relationship between
the mean value of $r_{02}$ and the {stellar} age.
This relationship was recently used by \citet{appourchaux2015} to 
estimate the age of the binary KIC~7510397 (HIP~93511).

Figure~\ref{fig:res_Age_r02} shows the distribution 
{of the mean of 
the  $r_{02}$ ratios, that is, $\langle r_{02} \rangle$,}
versus the derived ages for the sample of stars studied here.
A linear fit to these data leads to the following estimate of 
the stellar age, $\tau$ in Gyr, based on $ \langle r_{02} \rangle$
\begin{equation}
    \tau = 17.910 - 193.918 \langle r_{02} \rangle,
    \label{eqn:r02}
\end{equation}
{This is, of course, only valid for the range covered by our sample.
The range of radial orders used for calculating $\langle r_{02} \rangle$ has 
{almost no impact on
this result (an effect lower than a 1\%)}.
We note that when inserting the value of $\langle r_{02} \rangle = 0.068$ for the Sun,
Eq.~\ref{eqn:r02} yields an age of 4.7 Gyr, in excellent agreement with the Sun's
age as determined by other means.}

\section{Characterizing surface effects\label{sec:sec5}}

It is known that a direct comparison of observed frequencies
with model frequencies derived from 1D stellar structure models 
reveals a systematic discrepancy {that increases with 
the mode frequency}; this is commonly referred to as
\textup{\textup{{\it \textup{surface effects}} 
}}(\citealt{rosenthal1997}, see Section~\ref{sec1}).  
This discrepancy arises because a 1D stellar atmosphere does not represent the 
{actual structural and thermal properties of the 
stellar atmosphere in the layers close to the surface 
and because non-adiabatic effects that are present immediately below the surface are not
included when computing resonant frequencies using an adiabatic code.}
Some recent works {have attempted} to produce more realistic stellar atmospheres
by replacing the outer layers of a 1D stellar envelope by an 
averaged 3D surface simulation and by
including the effects of turbulent pressure in the 
equation of hydrostatic support and opacity changes from the 
temperature fluctuations, and by also 
considering non-adiabatic effects \citep{Trampedach2014,Trampedach2016,Houdek2016}.  
{This 
reduced the approximately $-$15~$\mu$Hz discrepancy to around 
+2~$\mu$Hz near 4,000 $\mu$Hz when including 
both structural and modal effects. }
While progress is being made, we are still not in a position to 
apply these calculations for a large sample of stars.

To sidestep this problem, several authors {have suggested} the use 
of combination frequencies that are insensitive to this systematic
offset in frequency, see for example, \citet{Roxburgh2003},
hence the exclusive use of $r_{01}$ and $r_{02}$ in
the AMP~1.3 method.
However, {since individual frequencies contain more information
than ratios of frequency separations,}
some authors have derived simple prescriptions to 
mitigate {the surface effects}. 
One such parametrization is that of \citet{Kjeldsen2008},
who suggested a simple correction to the 1D model frequencies 
$\delta\nu_{n,l}$ of the 
form {of a power law,}
\begin{equation}
    \delta\nu_{n,l} = a_0 \left ( \frac{\nu^{\rm obs}_{n,l}}{\nu_{\rm max}} \right )^{b}
        \label{eqn:kjeldsen}
,\end{equation}
where $b = 4.82$ is a fixed value, {calibrated by a solar model}, 
{ 
$\nu_{\rm max}$ is the frequency corresponding
to the highest amplitude mode, see \citep{Lund2016}}, 
$a_0$ is {computed from the differences between the} observed and model 
frequencies \citep{Metcalfe2009,Metcalfe2014}, 
\begin{equation}
a_0 = \frac{\langle \nu_{n,0}^{\rm obs} \rangle - \langle \nu_{n,0}^{\rm mod} \rangle}{N^{-1}_{0} \sum_{i=1}^{N_{0}} [\nu^{\rm obs}_i / \nu_{\rm max} ]^b}
.\end{equation}
{$\text{Here, }\nu^{\rm obs}_{n,l}$ and  $\nu^{\rm mod}_{n,l}$ are 
the observed and model frequency of radial order $n$ and 
degree $l$, respectively, 
and $N_{0}$ is the number of $l=0$ frequencies.}

In the absence of perfect 3D simulations, 
the interest in using such a surface correction becomes
evident when we consider {not only}
that the individual frequencies contain
a higher information content, but more importantly, {that} the
$r_{01}$ and $r_{02}$ frequency ratios
are only useful if the precision on these derived
quantities is high enough.
{A precision like this on the ratio requires not only having a high precision
on the individual frequencies, but
enough radial order modes to constrain
the stellar modeling.}  This is not necessarily the case for 
some stars, where, for example, ground-based campaigns are limited in time-domain
coverage, such as the case of $\nu$ Ind \citep{nuind}, 
or even for space-based missions such as the TESS mission, where 
{only one month of continuous data will be available for stars at 
certain galactic latitudes}.
{Similary, limited precision will also be achieved for the stars observed in the 
PLATO {\it \textup{step-and-stare}} phase, since the observation window
will only be two to three months each.}

The AMP~1.3 method exclusively uses the
$r_{01}$ and $r_{02}$ frequency ratios, and our results 
{are therefore expected to be} insensitive
to {surface effects}.  
{Hence}, using the resulting models and the observed frequencies,
we can explore the nature of the 
surface term for a large sample of stars, and in particular,
we can test to which extent the \citet{Kjeldsen2008} prescription is useful.

\subsection{Surface {effects} as observed in the Sun at low degrees}
\begin{figure}
    \begin{center}
    \includegraphics[width=0.48\textwidth]{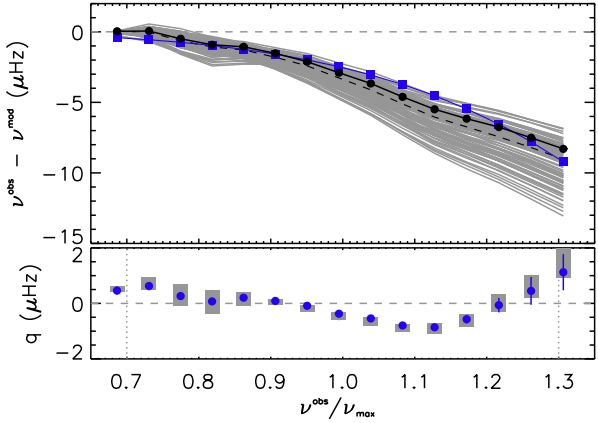}
    \caption{{\sl Top:} S{urface term} for the reference solar model 
    (connected black dots) for the $l=0$ frequencies as a function 
      of observed frequency scaled by \numax.
    The surface term for
    a sample of 100 of the best models is also shown for the Sun (gray), 
    with the mean
    value highlighted by the dashed line.  
    The blue connected squares show the empirical surface correction $\delta\nu_{n,l}$
    (Eq.~\ref{eqn:kjeldsen}) based on the reference model.
    {\sl Lower:} The differences between the observed and corrected model
    frequencies as a function of scaled frequency, with the solar observational
    errors overplotted in blue.  The shaded gray areas represent the mean and
    standard deviation of $q$ for the same 100 models shown in the top panel. 
    The dotted vertical lines delimit the region used to calculate
    the quality metric $Q$. }
    \label{fig:best100_surfaceterm}
        \end{center}
\end{figure}

{The magnitude of the surface {effects} on the frequency discrepancy for the Sun is on the 
order of 10-15 $\mu$Hz around 4000 $\mu$Hz for the low degrees ($l=0,1,2,\text{and }3$)}.
Our
analysis using the solar data reveals a similar offset.
In the top panel of Fig.~\ref{fig:best100_surfaceterm} we show the
solar surface term by comparing the input frequencies
with those of the models. 
The term of the reference model  is 
shown by the thick line with filled black dots, and in 
gray we show those  
for 100 of the best solar models, with the mean of these 100 shown as the thick dashed line.
At \numax, the value of $a_0 = -2.5 \mu$Hz for the reference model,
and for 100 of the representative models it spans --2.3 to --4.6 $\mu$Hz.

When we apply Eq.~\ref{eqn:kjeldsen} to the reference solar model, we calculate
a correction $\delta\nu_{n,l}$ that successfully mitigates the surface {effects}.
This is clearly shown in the top panel of Fig.~\ref{fig:best100_surfaceterm} ,
where the surface {term} for the reference model (black connected dots) is traced by the
scaled surface correction $\delta\nu$ (blue connected squares) for the 
$l=0$ modes alone.  
By applying the proposed corrections $\delta\nu_{n,l}$ to the 
observed frequencies, we can then make a quantitive
comparison between the model and the data. 
This agreement is shown in the lower panel for $l=0,$ and we denote it as 
    $q_{n,l} = \nu^{\rm obs}_{n,l} - \nu^{\rm mod}_{n,l} + \delta\nu_{n,l}$.
To quantify the agreement between the corrected model frequencies and the observed ones, 
we define the metric $Q$ as the median of the 
{absolute value} of the residuals,
\begin{equation}
 \\
   Q = \mathrm{median}\left | {q_{n,l}} \right |, 
    \label{eqn:chit}
\end{equation}
for all observed $n$ and $l$ defined in the region of 
$0.7 \le \nu_{n,l}^{\rm obs} / \nu_{\rm max} \le 1.3$.
This region is delimited in the lower panel by the vertical dotted lines.
We note that we purposely exclude any reference to an observational 
error in the 
definition of $q$,
as the surface correction results from an error in the models and is not related to the 
precision of the frequency data.  
In the ideal case and in the absence of errors in the data, $Q \rightarrow 0~\mu$Hz, which means that the model is perfect.
The value of 
$Q$ is 0.38 $\mu$Hz for the reference solar model, and the mean
value for the 100 solar models shown in 
Fig.~\ref{fig:best100_surfaceterm} is 0.51 $\mu$Hz. 
From this figure and the low value of the quality metric, it is expected that the 
\citet{Kjeldsen2008} empirical surface correction $\delta\nu_{n,l}$ (Eq.~\ref{eqn:kjeldsen})
is useful for mitigating {the surface effects} for this solar model.

\subsection{{Surface effects} for other stars}

Is the simplified surface correction useful in other stars? And 
if so, to what extent?  
These are the questions that we aim to answer by inspecting the 
reference models (Table~\ref{tab:referencemodels})
of the best-fit stars within our sample.

\begin{figure}
    \begin{center}
    \includegraphics[width=0.48\textwidth]{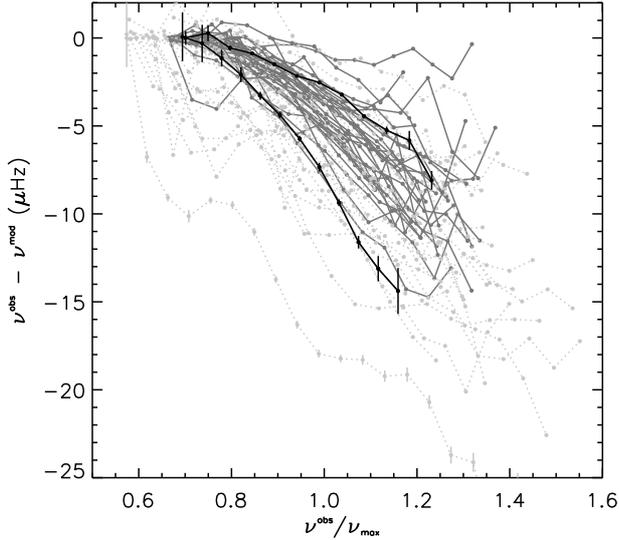}
    \caption{Surface terms for the stars in our subsample defined by
      the criteria of $\chi_{N}^2 (r_{01},r_{02}) \le 3$.
      For clarity, the more evolved
    stars are shown by the dashed lines.}
    \label{fig:star_surfacecorr}
        \end{center}
\end{figure}

We define a subset of stars by selecting those with
$\chi^2_N \leq 3.0$\footnote{The limit of 3.0 is rather arbitrary
and was chosen as a compromise between having an adequate sample size
and the best match to the data.
Using a threshold of 2.0 or 4.0 does not change the results significantly.} for both $r_{01}$ and 
$r_{02}$. 
This {selection results in a subset of} 44 stars.
The differences between the observed frequencies and 
the frequencies of the reference models for this subset
are shown in Fig.~\ref{fig:star_surfacecorr}, 
{and we assume} that these differences are dominated 
by {the surface effects.}
For the stars represented by the continuous lines it can be noted
that the {remaining discrepancies are} quite similar in 
{magnitude} and shape for the {less evolved stars}.
{For} the more evolved stars {(\logg\ $>$ 4.2, indicated
by dashed lines), the remaining discrepancies are larger and {of a different nature},
and cannot readily be modeled by a simple power law.}

\begin{figure}
    \begin{center}
    \includegraphics[width=0.48\textwidth]{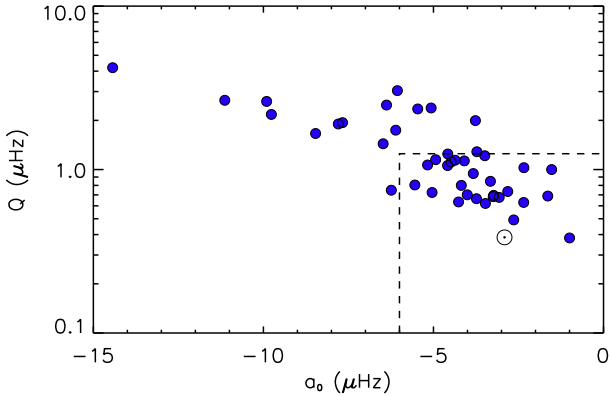}
    \caption{Metric $Q$ versus $a_0$ for the stars in our subsample.
    {The dashed lines highlight the approximate limitation in $Q$ and $a_0$ 
    where the surface correction enables a useful comparison
    between the observed and corrected model frequencies}.}
    \label{fig:dif_cut}
        \end{center}
\end{figure}

For each of the stars, {a value of $a_0$ is derived directly from 
the comparison of model and observed frequencies (see Table~\ref{tab:properties_derived}),}
and 
Eq.~\ref{eqn:kjeldsen} is used to calculate the surface
correction $\delta\nu_{n,l}$ to apply to the model frequencies.
We then calculate the metric $Q$ for each star in the subsample,
{and these values} are shown as a function of $a_0$ in
Fig.~\ref{fig:dif_cut}.
We see very clearly that as the difference between the 
observed and model frequency at \numax\ increases (i.e., $a_0$ becomes more negative), $Q$ also increases,
indicating that the \citet{Kjeldsen2008} correction becomes less {adequate} 
to mitigate the surface {effects}.  
It seems then quite likely that there is a value of $Q$ (and $a_0$) that defines
a limit where the surface correction is useful.

By inspecting the residuals between observed and corrected model frequencies 
for this subset of stars, we 
found that when $Q \lesssim 1.0~\mu$Hz, we obtained a very good 
match to the observed frequencies when the surface correction was included.
These stars also have values of $a_0$ that are typically lower than {$-$6.0~$\mu$Hz}, as shown in
Fig.~\ref{fig:dif_cut}, just like the solar case.  
For {an illustration}, we present some \'echelle diagrams 
in Fig.~\ref{fig:someechelles} with different values of $Q$
to {show} the validity of this criterion.
{A visual inspection of the residuals and the \'echelle diagrams
for this subsample of stars led to the same conclusion.}

\begin{figure*}
    \centering
\includegraphics[width=0.48\textwidth]{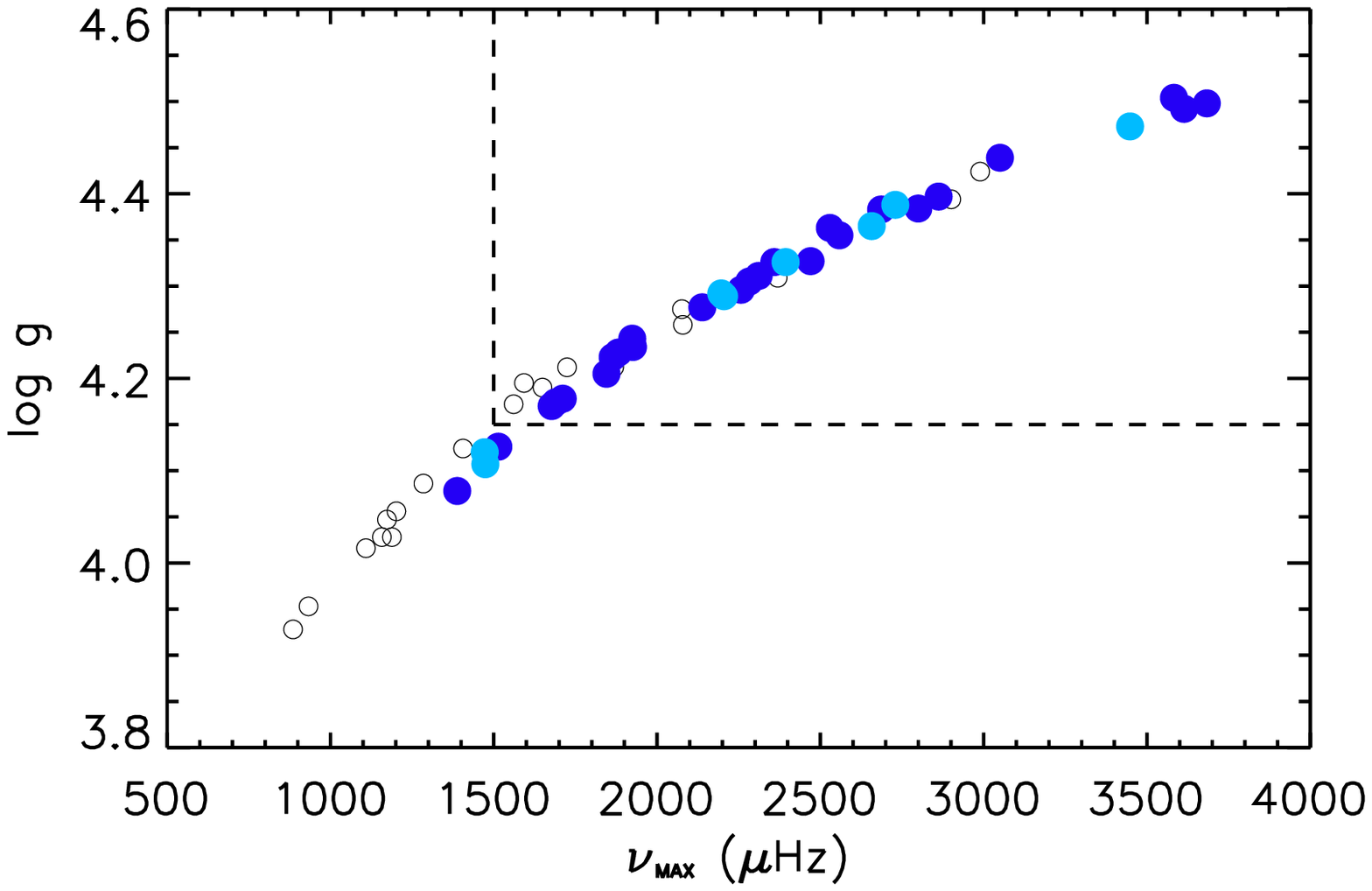}
    \includegraphics[width=0.48\textwidth]{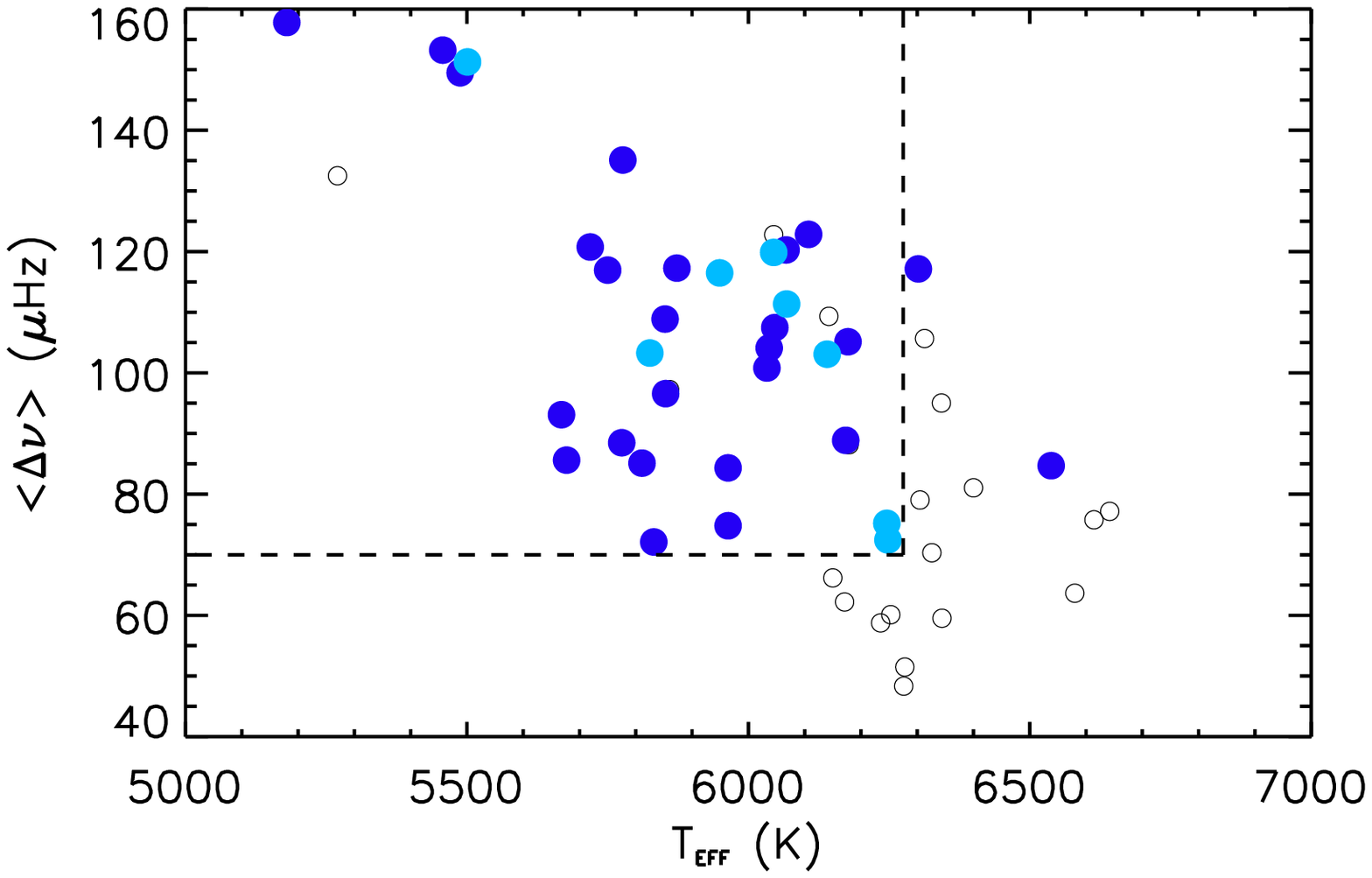}
    \includegraphics[width=0.48\textwidth]{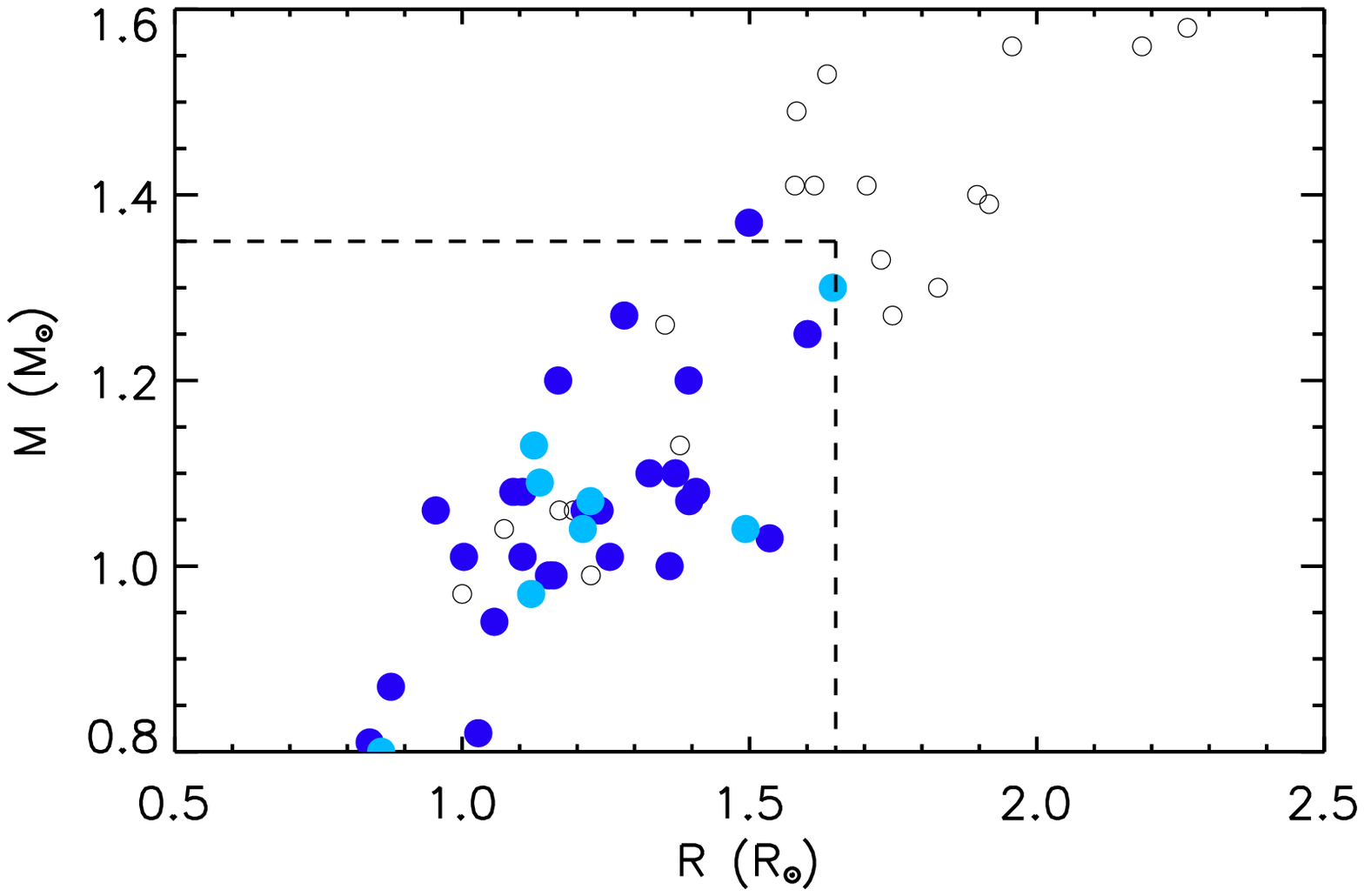}
     \includegraphics[width=0.48\textwidth]{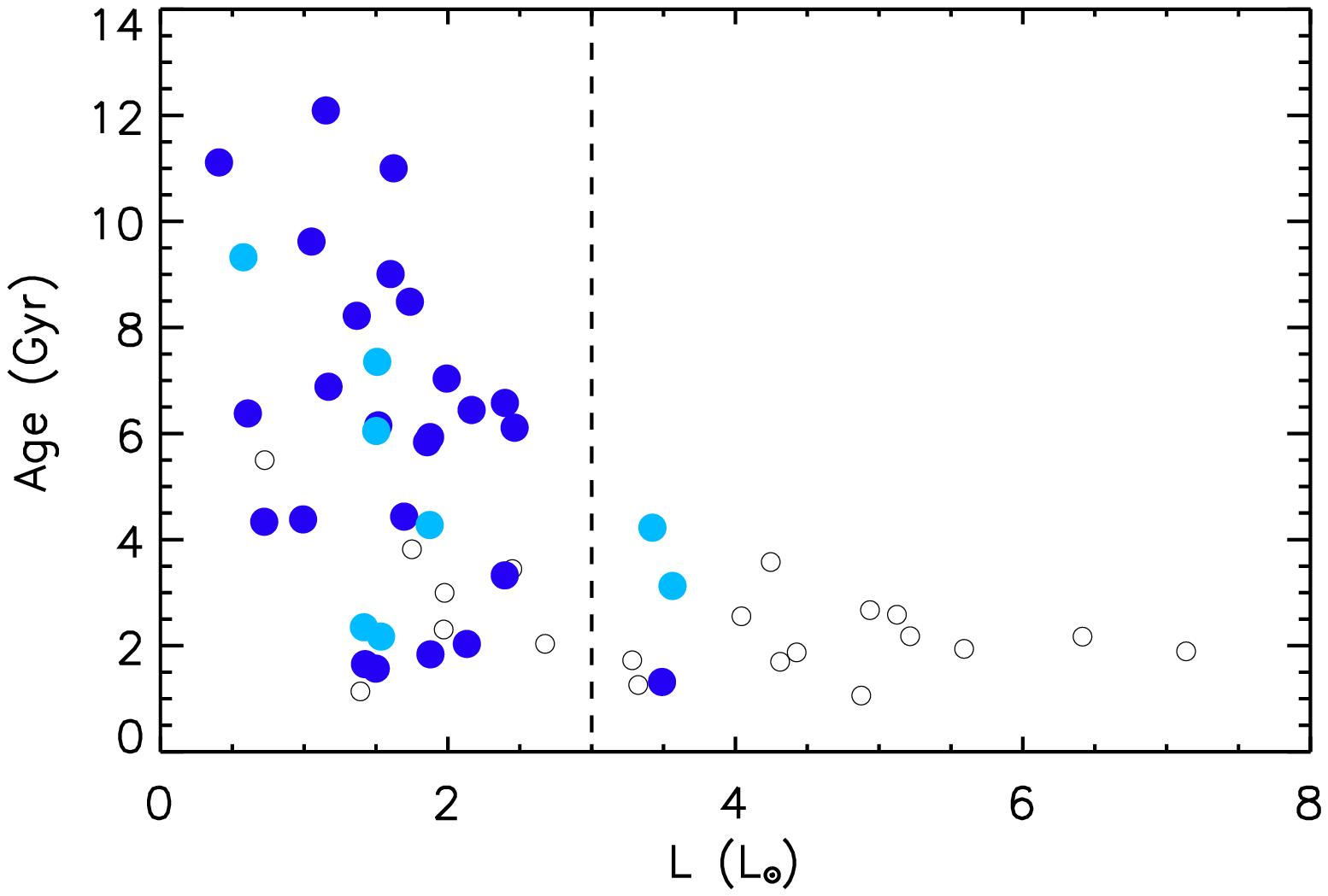}
\caption{Distribution of observed (top panels) and derived 
parameters (lower panels) for the {selected subsample of stars (open circles).
The dark and light blue filled dots represent the stars with $Q \le 1.0$ and 1.2.
The regions are delimited by dashed lines within which we infer
that the 
\citet{Kjeldsen2008} surface prescription should be useful.}}
    \label{fig:whichparameterswork0}
\end{figure*}

When we rely on the criteria of $Q \lesssim 1$ $\mu$Hz, we can trace the 
ranges of the stellar parameters where the surface correction 
 mitigates the surface {effects}.  This is {presented} in 
Fig.~\ref{fig:whichparameterswork0}, {which shows} 
the distribution of 
observed and inferred stellar properties of stars from this subsample (open circles)
along with the stars that satisfy the criterion of 
$Q \lesssim 1.0~\mu$Hz (filled dark blue circles) and
$Q \lesssim 1.2~\mu$Hz (filled light blue circles).  
We also delimit the regions (dashed lines) where we infer that
the correction is no longer useful. 

More concretely, we find that the limit of the solar-like regime
in terms of observed properties is 
approximately at $\log g = 4.2$, \teff\ $= 6250$ K, 
\mlsep\ $= 70 \mu$Hz and \numax\ $= 1600 \mu$Hz.
In terms of physical properties of the star, the limit is 
around $R = 1.6$ \rsol, $M = 1.35$ \msol, and $L = 3.0$ \lsol,
with no evidence that the absolute age (not evolution state) 
or the metallicity
playing any role.  
In Table~\ref{tab:appliedsurface} we summarize these 
{limiting regions, but adopt a slightly} more conservative limit.

\begin{figure}
    \centering
\includegraphics[width=0.48\textwidth]{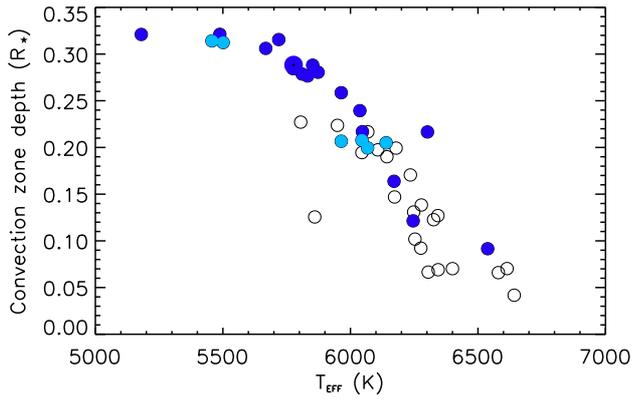}
 \caption{Fractional depth of the convection zone as a function of \teff\ for our
selected subsample of stars.  The color-coding is 
the same as Fig.~\ref{fig:whichparameterswork0}.}   \label{fig:teffdcz}
\end{figure}

The limit in \teff\ can probably be attributed to the Kraft break (e.g., \citealt{Kraft1967}), where
at around $6250$ K, these hotter stars rotate much faster 
as a result of a lack of a deep convective envelope, 
in which magnetic braking could slow the star down. 
The depth of the convective region is shown as a function 
of \teff\ in Fig.~\ref{fig:teffdcz}, and stars with regions larger than approximately
0.2 stellar radii satisfy this criterion.  This limit is also compatible with the proposed mass 
limit of approximately 1.3~\msol\ where a transition in envelope convection 
takes place.  
The negative slope of the surface correction at \numax\ is also
found to increase with increasing mass (becoming flatter), again
indicating a change in convective zone properties and \teff.
The limit in \logg\ points toward a transition 
from the main-sequence to the subgiant phase where the 
convective envelope begins to deepen.

These limits are imposed by the physical structure of the star itself,
but no quantitative measure of $a_0$ can be deduced from the 
observed and/or inferred stellar properties at this stage, except for 
a slight linear dependence of $a_0$ with \mlsep, \numax, or \logg\ with
a rather large scatter.

\begin{table}
\begin{center}
\caption{Stellar property regimes where the \citet{Kjeldsen2008} surface correction is useful.
\label{tab:appliedsurface} }
\begin{tabular}{rlllllll}
\hline\hline
Property & \multicolumn{2}{l}{Value} \\
\hline
\logg\ (cgs) & $\ge$ & 4.2 \\
\teff\ (K)& $\le$ & 6200 \\
\mlsep\ (\mhz)&$\ge$& 80 \\ 
\numax\ (\mhz)& $\ge$& 1700 \\
$a_0$ (\mhz)& $\le$& -6 \\
$R$ (\rsol)& $\le$& 1.5\\
$M$ (\msol)& $\le$& 1.3\\
$L$ (\lsol)& $\le$& 2.5\\
\hline\hline
\end{tabular}
\end{center}
\end{table}

\section{Summary\label{sec6}}

The high-quality and long-term photometric time series
provided by {\it Kepler} has enabled an unprecedented
precision on asteroseismic data
of stars like the Sun.  
Thanks to the very high precision, we could {use} the frequency separation
ratios along with spectroscopic temperatures and metallicities
to infer stellar properties of the Sun and 57 {\it Kepler} stars, 
comprising solar analogs, active stars, 
components of binaries, and
planetary hosts, with a precision of the same quality
when using the individual frequencies. 
Median uncertainties on radius and mass are 1\% and 3\%,
while uncertainties on the age compared to the estimated
main-sequence lifetime are typically 7\% 
or 11\% compared to 
the absolute age.  
These realistic uncertainties {account for
unbiased determinations of mixing-length
parameter and initial chemical composition}.
Along with the physical stellar properties, we also derived the 
interstellar absorption and distances to each star, and where 
the rotation period was available, we derived the rotational velocity. 
For nine stars {our derivation of radii, luminosities, and distances
are in very good agreement with independently measured values.} 
{Our inferred ages are} validated
for the Sun and by comparing the ages of the individual components
of the binary system 16 Cyg A and B. 

From an analysis of our derived properties for the full sample 
we investigated the {dependence} of the mixing-length parameter
with stellar properties 
and found it to correlate with \logg\ and 
\teff\ , just as proposed by \citet{magic2015} from 3D RHD simulations
of convective envelopes.  
We also derived a linear expression relating the mean value of
the $r_{02}$ frequency separation ratios directly to the age of the star, which
yields an age of 4.7 Gyr for the Sun. 

By selecting a subsample of the stars using a $\chi^2_N$ {threshold},
we investigated the usefulness of the \citet{Kjeldsen2008} 
empirical {correction for the surface effects} across a broad range of stellar parameters, 
and we found that it is useful, {but only} in certain regimes, 
{as also suggested by the theoretical study of \citet{schmittbasu2015}}.  
This is of particular interest for stars with 
much shorter time series, where the precision on the 
individual frequencies or the number of radial orders
is not high enough  
to constrain the stellar modeling.
In particular, this
will be the case for the forthcoming NASA TESS mission, 
where some stars with ecliptic latitude $|b| \lesssim 60^{\circ}$ will 
be observed continuously for only 27 days, 
along with the {\it \textup{step-and-stare}} 
phase of the future PLATO mission (launch 2024). 

\section{Perspectives\label{sec7}}

In this work we used  
\teff\ and \mh\ as the only complementary data to the asteroseismic 
data.  However, within a year from now, we will have 
a homogenous set of microarcsecond precision parallaxes that
will give access to the intrinsic luminosity of the star.
{This quantity is sensitive to the interior stellar composition.} 
While today we have very high precision radii along with other properties,
degeneracies in model parameters, such as  the mass and initial
helium abundance (e.g., \citealt{Metcalfe2009,lebreton2014})
limit the full exploitation of 
asteroseismic data for testing stellar interior models and 
improving precision on model parameters.  
The forthcoming Gaia data in Release 2 promise to 
overcome this obstacle 
and thus provide 
even higher precision radii and ages, along with constraints on 
interior and initial chemical composition,  and thus pushing stellar models
to their limit. 

\begin{figure}
    \includegraphics[width=0.48\textwidth]{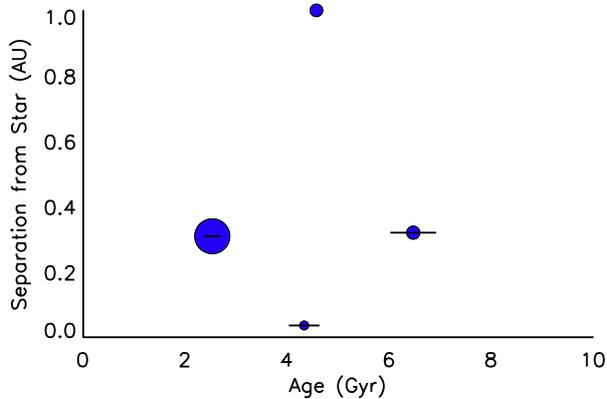}
    \caption{Age of planet and separation from host.  Symbol sizes represent planetary radius, 
    and equilibrium temperature decreases with distance from the host.
    Age and radius are taken from this work, while other parameters are taken from \citep{batalha2013}. The Earth is shown at 1 AU.
    \label{fig:planetages}}
\end{figure}

We highlight the importance of the precise characterisation of exoplanetary
systems using asteroseismic data. In this work, we determined the radius and age 
of three planetary hosts (KIC~9414417, KIC~9955598, and KIC~10963065).  
Combining our data with 
those of \citet{batalha2013} constrains the 
planetary and orbital parameters.  We illustrate this in 
Fig.~\ref{fig:planetages}, where we depict the separation of the 
planet and host as a function of stellar age (including the Earth).  
The sizes of the symbols {are indicative of} the planetary radius, and 
the equilibrium temperature decreases with distance from the host.  
The diversity of planetary systems can be easily noted, and 
such an analysis of a larger sample of planetary candidates
will yield important constraints on the formation and evolution of planetary 
systems.  
The future TESS and PLATO missions targeting bright stars with 
asteroseismic characterization promise to be a goldmine for not only
exoplanetary physics, but with access to microarcsecond parallaxes 
and homogenous 
multiband photometry, also for stellar and Galactic physics.

\begin{acknowledgements}
This work is based on data collected by the Kepler mission. Funding for the Kepler mission is provided by the NASA Science Mission directorate.
This collaboration was partially supported by funding from the Laboratoire Lagrange 2015 BQR.
This research has made use of the VizieR catalogue access tool, CDS, Strasbourg, France. The original description of the VizieR service was published in A\&AS 143, 23.
This work was supported in part by NASA grants NNX13AE91G and NNX16AB97G. Computational time at the Texas Advanced Computing Center was provided through XSEDE allocation TG-AST090107. 
DS and RAG acknowledge the financial support from the CNES GOLF grants.  
DS acknowledges the Observatoire de la C\^ote d'Azur for support during his stays.  Some of these computations have been done on the ’Mesocentre SIGAMM’ machine, hosted by the Observatoire de la Cote d’Azur.
The authors wish to thank Sylvain Korzennik for his very careful reading of the 
paper and valuable suggestions for improving the presentation and the scientific arguments.
\end{acknowledgements}

\begin{appendix}

\section{Supplementary material\label{sec:app0}}

\begin{figure*}
    \centering
    \includegraphics[width=0.48\textwidth]{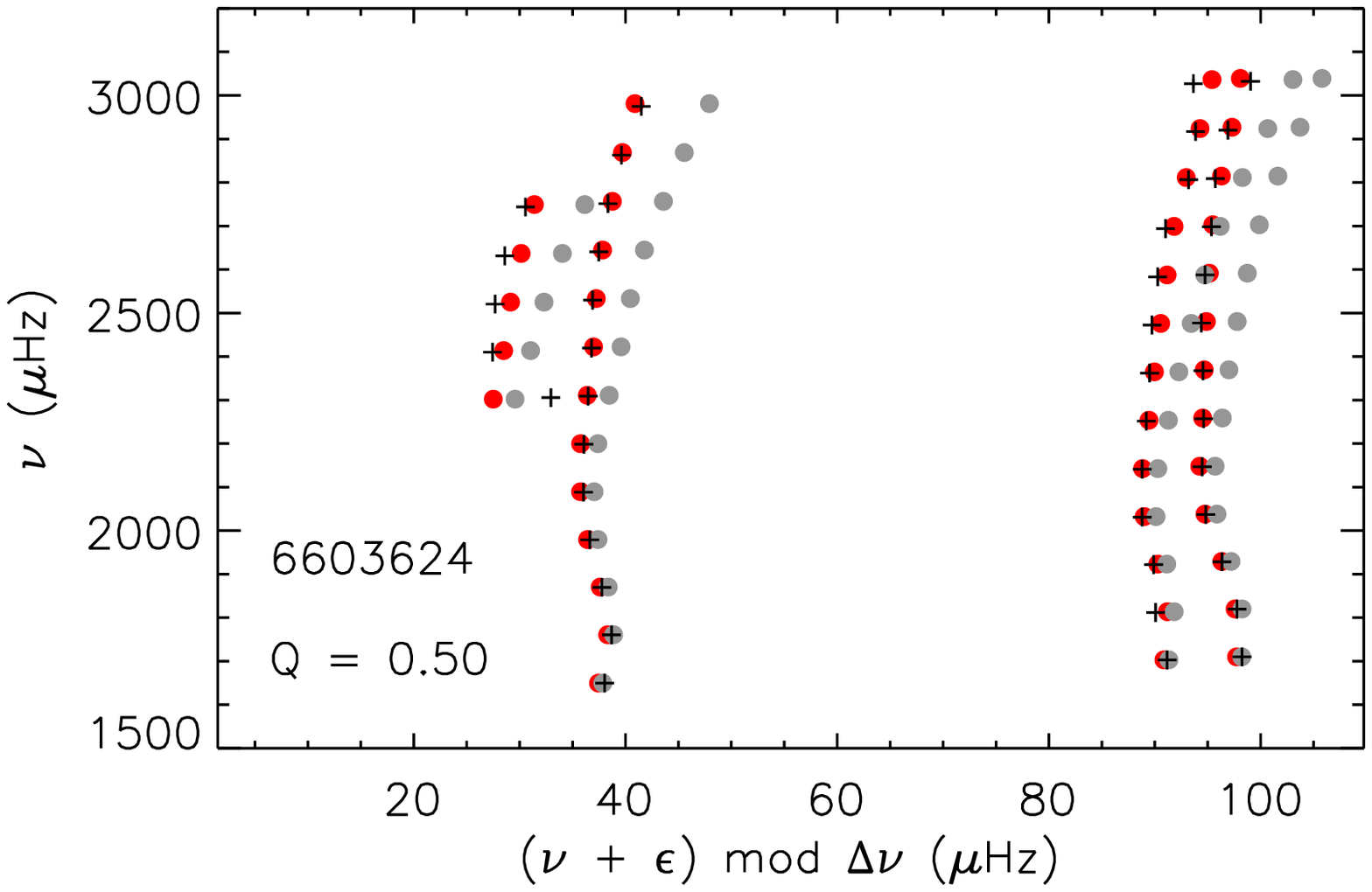}
    \includegraphics[width=0.48\textwidth]{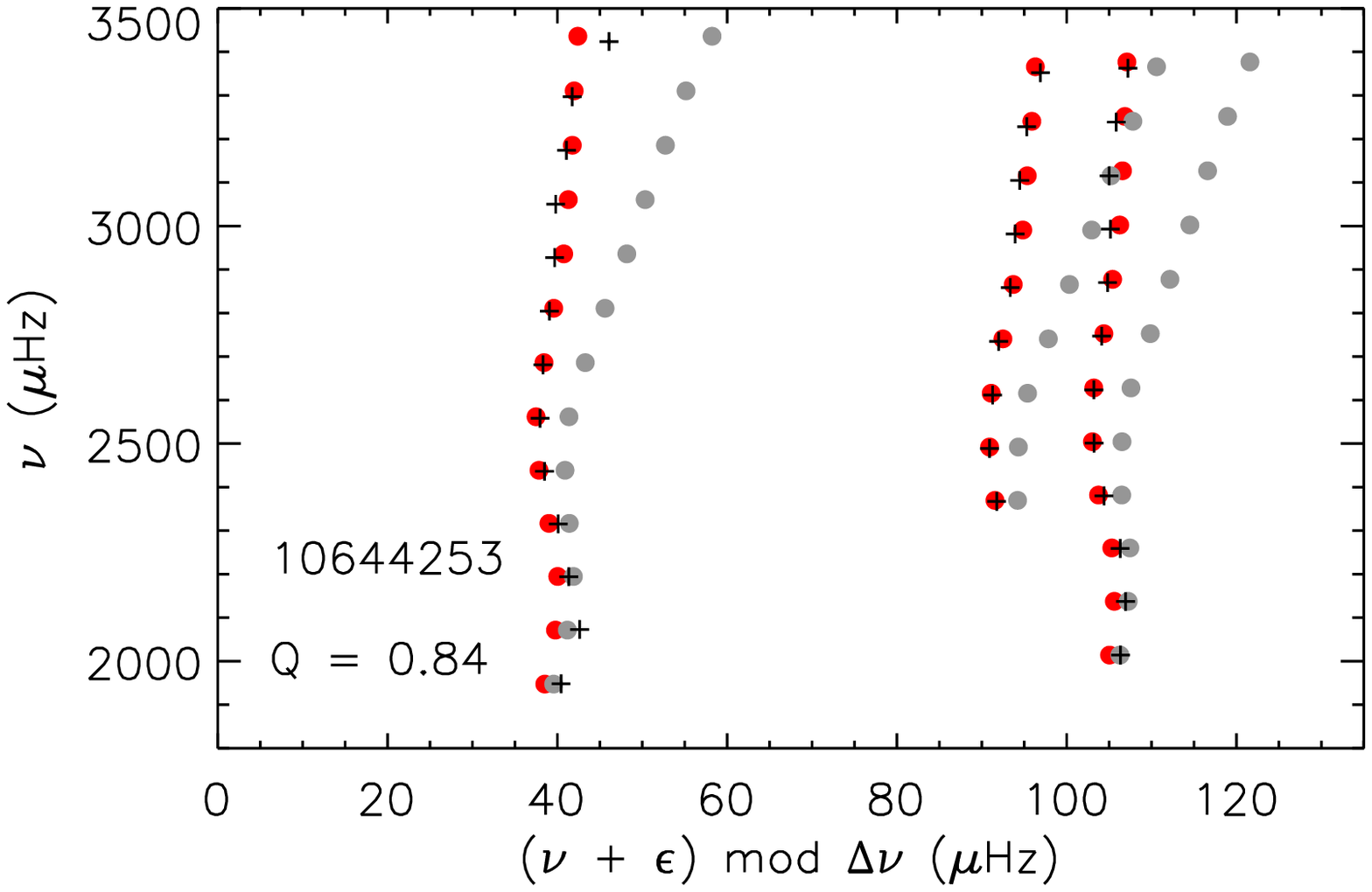}
    \includegraphics[width=0.48\textwidth]{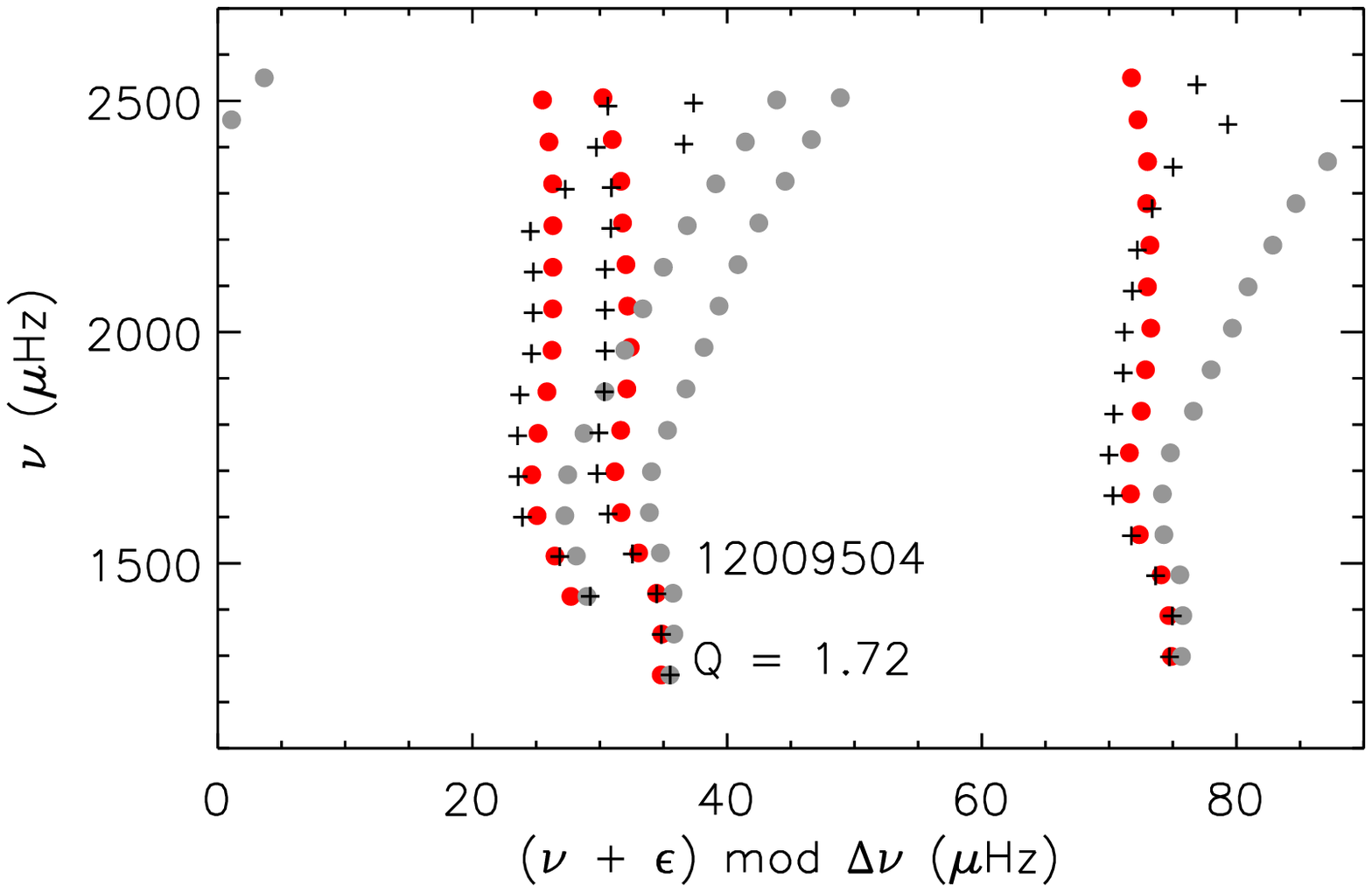}
    \includegraphics[width=0.48\textwidth]{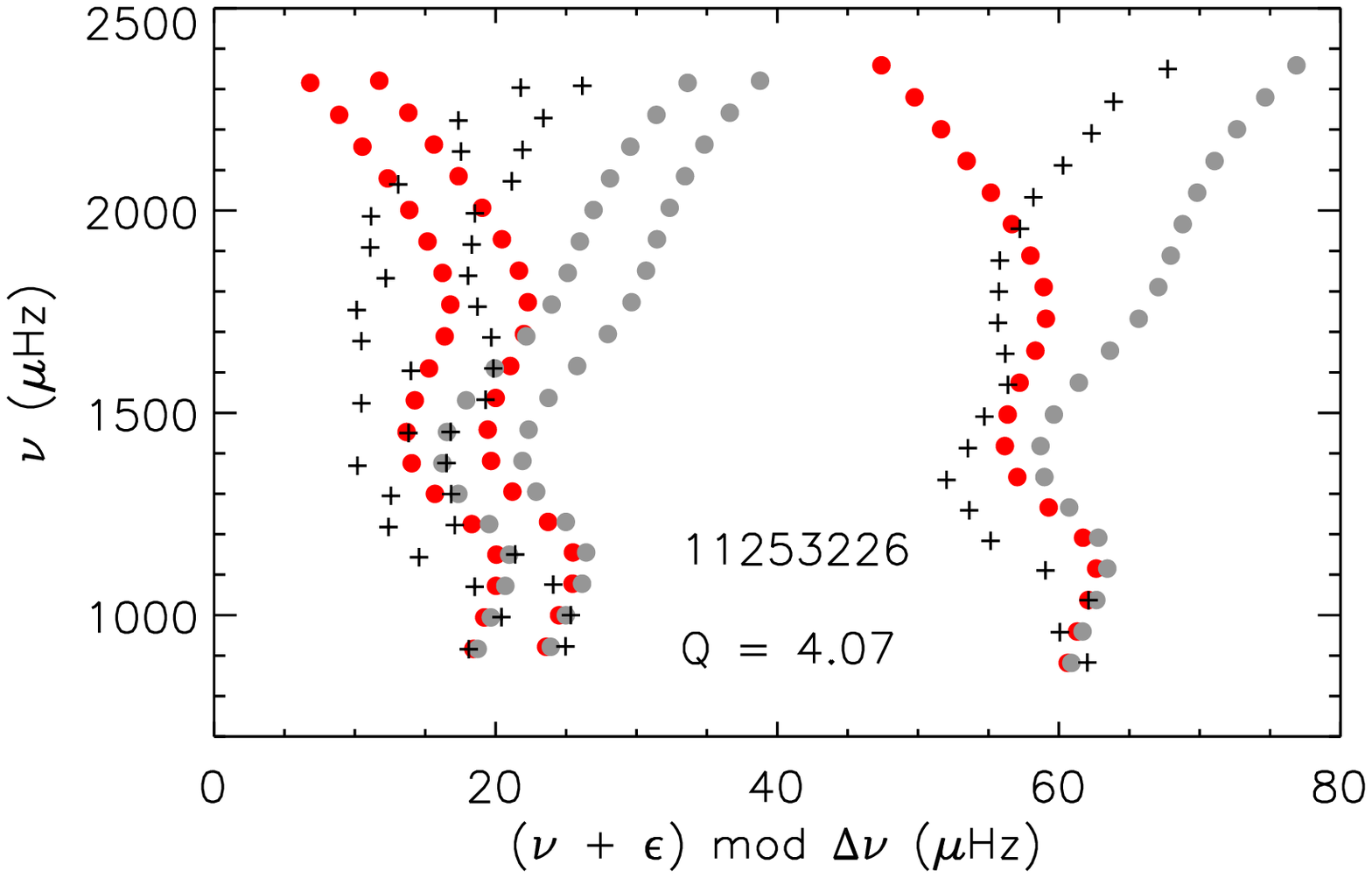}
    \caption{Echelle diagrams for two stars where the 
    surface correction appears to be useful (top) and for two stars where
    the correction is not useful (lower). 
Crosses are observed frequencies, and black circles
    are corrected frequencies. The value of $\epsilon$ is an
    arbitrary shift in x-axis for display purposes.}
    \label{fig:someechelles}
\end{figure*}

\begin{table*}[h!]
\begin{center}\caption{Derived stellar properties of {\it Kepler} targets with M $\sim$ 1.2 
\msol\ with or without diffusion of helium \label{tab:diffusioneffects}}
\begin{tabular}{lccccccccc}
\hline\hline
KIC ID & $R$ & $M$ & Age & $L$ & $\log g$ & [M/H] \\
& (\rsol) & (\msol) & (Gyr) & (\lsol) & (dex) & (dex)\\
\hline
no diffusion\\
9139151   &  1.132  &    1.11  &    1.96  &    1.80  &    4.375&  -0.01  \\
12009504  &  1.366  &    1.10  &    3.38  &    2.39 &     4.210&   -0.01\\
6225718  &   1.227  &    1.15  &    2.29  &    2.09  &    4.320&    -0.10\\

\hline
diffusion\\
1225814 &  1.595  &  1.26  &    5.04  &   2.81  &  4.129 &  0.05\\
5184732 &  1.356  &   1.25 &    4.68 &    1.82 &   4.269&   0.25\\
8150065 &  1.397  &   1.21 &    3.12 &    2.54 &   4.228& -0.05\\
8179536 &  1.348  &   1.25 &    1.93 &    2.64 &   4.274&  -0.05\\
7771282 &  1.631  &   1.26 &    3.34 &    3.65 &   4.116&  -0.03\\
10454113&  1.250  &   1.20 &    1.98 &    2.04 &   4.320&  -0.04\\
\hline\hline
\end{tabular}
\end{center}
\end{table*}

\end{appendix}

\bibliographystyle{aa}
\bibliography{seismic}

\end{document}